%% file: 0_Main_manuscript_revised.tex
\documentclass[gmd, manuscript]{copernicus}

\usepackage{tabularx}
\usepackage{cancel}
\usepackage{multirow}
\usepackage{supertabular}
\usepackage{algorithmic}
\usepackage{algorithm}
\usepackage{amsthm}
\usepackage{float}
\usepackage{subfig}
\usepackage{rotating}

\usepackage{longtable}
\usepackage{hyperref}
\usepackage{amsmath, amssymb}
\nolinenumbers
\begin{document}

\title{A PMP-inspired Evaluation Framework for Assessing
% the fit-for-purpose of % can be taken out (suitability)
Deep-Learning Earth System Models}
% A PMP-guided framework for evaluating the reliability of AI-driven weather models
% \Author[1]{Multiple Authors}{}

\Author[1,*]{Giuliana}{Pallotta}
\Author[1,*,\textdagger]{Shiheng}{Duan}
\Author[1]{Céline}{Bonfils}
\Author[1]{Jiwoo}{Lee}
\Author[1]{Seth}{Goodnight}
\Author[1,2]{Paul}{Ullrich}

% \Author[Giuliana Pallotta][pallottagold1@llnl.gov]{}{} %% correspondence author
\affil[1]{Lawrence Livermore National Laboratory}
% \affil[2]{Gridmatic, formerly at Lawrence Livermore National Laboratory}
\affil[2]{University of California Davis}
% \affil[]{ADDRESS}
\affil[*]{These authors contributed equally to this work.}
\affil[\textdagger]{Now at Gridmatic}
%% The [] brackets identify the author with the corresponding affiliation. 1, 2, 3, etc. should be inserted.

%% If an author is deceased, please add \deceased[$Deceased date if applicable$]{$Author number$} (e.g. \deceased[13 November 2015]{2}) at the end of the affiliations. The author number depends on the placement of the author in the author list, e.g. the third author has number 3.

%% If authors contributed equally, please add \equalcontrib{$Author numbers$} (e.g. \equalcontrib{1,3}) at the end of the affiliations. The author number depends on the placement of the author in the author list, e.g. the third author has number 3.

\runningtitle{PMP for DL-ESMs}

\runningauthor{TEXT}

\received{}
\pubdiscuss{} %% only important for two-stage journals
\revised{}
\accepted{}
\published{}

%% These dates will be inserted by Copernicus Publications during the typesetting process.

\firstpage{1}

\maketitle

% Limit to 250 words
\begin{abstract}
In recent years, Deep-Learning Earth System Models (DL-ESMs) have emerged as promising, computationally efficient complements to traditional Earth system models.
Here, we present an evaluation framework for testing DL-ESMs from an Earth system model-development perspective using standardized diagnostics from the PCMDI Metrics Package (PMP).
This framework allows DL-ESMs, including Ai2's ACE2 and Google's NeuralGCM, to be assessed with metrics that quantify their ability to reproduce climatology, major modes of variability, monsoons, and precipitation variability relative to observational reference datasets and CMIP-class benchmarks.
By evaluating DL-ESMs with tools commonly used for traditional models, we extend their assessment beyond short-range forecast skill and toward longer Earth System-relevant applications.
The results identify encouraging strengths in several large-scale fields and modes of variability, while also highlighting persistent challenges in precipitation, tropical variability, and long-run stability for some model versions.
This evaluation is a critical step toward building trust in DL-ESMs, guiding future model development, and clarifying their fit-for-purpose for Earth system science applications.
\end{abstract}
%%%%%%%%%%
% \copyrightstatement{TEXT} %% This section is optional and can be used for copyright transfers.

\introduction
Earth System Models (ESMs) are essential tools for understanding past, present, and future Earth System variability and extremes. By coupling representations of the atmosphere, ocean, land surface, and cryosphere, ESMs provide a physically consistent framework to better understand the different processes governing the Earth system and to support a wide range of applications. For instance, ESMs support seasonal-to-decadal prediction, enable the simulation and attribution of extreme events \citep{smith2019robust}, and provide a basis for building physically plausible storylines in which observed extremes are replayed under altered thermodynamic conditions \citep{shepherd2018storylines}. ESMs are also used in nudging experiments that isolate the drivers of particular events \citep{Pithan2023Nudge}, underpin global and regional future projections and past climates within the framework of the Coupled Model Intercomparison Project (CMIP) \citep{Zhang2025fut}.
% , and constrain climate sensitivity \citep{Coats2020,Kageyama2024paleo}.
% However, many of these ESM-based applications remain computationally expensive, particularly when they require high-resolution simulations, long integrations, or large ensembles, motivating interest in Deep Learning (DL) and Artificial Intelligence (AI) approaches that can complement traditional modeling by accelerating selected model components, emulating costly processes, supporting large-ensemble generation, improving downscaling and post-processing, and extracting information from increasingly large Earth system datasets.
These applications, however, place substantial demands on computational resources: they often require high spatial resolution, long integrations, large initial-condition or perturbed-physics ensembles, repeated sensitivity experiments, and extensive post-processing. This computational bottleneck has motivated growing interest in Deep Learning (DL) and Artificial Intelligence (AI) as complementary tools for Earth science, including the emulation of costly physical processes, acceleration of selected model components, support for ensemble generation, improvement of downscaling and post-processing workflows, and extraction of information from increasingly large observational and model archives.
% Recent advances in Deep Learning (DL) and Artificial Intelligence (AI) have accelerated progress in weather forecasting science. Deep-Learning Weather Prediction Models (DL-WPMs), such as FourCastNet \citep{FourCastNet}, Pangu-Weather \citep{Pangu-weather}, and GraphCast \citep{Graphcast_2023}, have demonstrated that data-driven systems can generate skillful atmospheric forecasts at substantially lower computational cost than many traditional numerical weather-prediction workflows. Unlike physics-based numerical weather prediction models, which explicitly solve discretized governing equations on large supercomputers, DL-WPMs learn atmospheric evolution from large historical datasets and then advance the state autoregressively.
%%%%%%%%%%%%%%
This interest has been particularly visible in weather forecasting, where recent advances in DL and AI have accelerated progress. Deep-Learning Weather Prediction Models (DL-WPMs), such as FourCastNet \citep{FourCastNet}, Pangu-Weather \citep{Pangu-weather}, and GraphCast \citep{Graphcast_2023}, have demonstrated that data-driven systems can generate skillful atmospheric forecasts at substantially lower computational cost than many traditional numerical weather-prediction workflows. Unlike physics-based numerical weather prediction models, which explicitly solve discretized governing equations on large supercomputers, DL-WPMs learn atmospheric evolution from large historical datasets and then advance the state autoregressively.
These successes have naturally raised the question of whether similar data-driven approaches can be extended from short- to medium-range weather prediction to the broader and more demanding problem of Earth system simulation.

Deep-Learning Earth System Models (DL-ESMs) are now emerging as fast surrogate or hybrid alternatives for Earth System simulation. However, Earth System applications require more than simply running a weather-prediction model for longer lead times: DL-ESMs must remain stable over extended roll-outs, respond plausibly to prescribed boundary conditions and external forcings, and reproduce not only the mean climatology but also key modes of variability and extremes. Current DL-ESMs therefore represent a potentially transformational pathway by combining initial-condition information with boundary-condition constraints such as insolation, sea-surface temperature (SST), sea-ice concentration (SIC), and fixed forcings \citep{Eyring2024, Camps2025, Price2025}. Their promise is substantial, but they also pose distinct evaluation challenges, including conservation of mass and energy, physical consistency, long-term stability, and generalization outside the training distribution.
Examples of this rapidly developing class of DL-ESMs include Ai2's ACE2 Earth System emulator \citep{wattmeyer2023,watt2025ace2}, which uses deep learning to emulate atmospheric behavior at a fraction of the cost of full-fidelity ESMs, and recent ACE2 extensions that explore coupling to a slab ocean model \citep{ACE2-SOM_2025}. Google's NeuralGCM \citep{Kochkov2024} follows a hybrid approach by combining a differentiable dynamical core with learned parameterizations; recent variants have added enhanced precipitation capabilities \citep{Yuval2026}. Another recent example, DLESyM \citep{DLESyM_2025}, couples an atmospheric U-Net model with a SST ocean model and has shown promise for studying internal Earth System variability and subseasonal-to-seasonal prediction, including recent encouraging results for extrapolation \citep{Meng2026}.

The growing availability of DL-ESMs creates an urgent need for evaluation strategies that are comparable to those used for traditional ESMs. While standard weather-forecast scores are informative, they remain insufficient for Earth system studies, because a model that performs well at short lead times may still misrepresent mean-state biases, regional hydrological processes, low-frequency variability, or long-term stability. The long history of ESM evaluation therefore provides an important road map for assessing DL-ESMs, especially within historical periods where common observational references and AMIP-style boundary conditions are available \citep{Ullrich2025}.

Several recent studies have begun to address this need. \citet{Zhang2025} compared ACE2, NeuralGCM, and cBottle, a generative diffusion model developed by NVIDIA \citep{Brenowitz2025ClimateBottle}, in an out-of-sample uniform SST warming experiment benchmarked against GFDL's AM4 \citep{GFDLAM4}. Their results showed that the tested models retain some key responses, including aspects of precipitation, but struggle with extrapolation beyond the training distribution. \citet{Rucker2025bench} evaluated regional thermodynamic trends in ACE2 and NeuralGCM using ERA5 boundary conditions with AMIP simulations, finding that ACE2 captures near-surface warming trends in the mid-latitude troposphere more effectively than NeuralGCM, whereas both DL-ESMs outperform physics-based models in representing Arctic amplification. \citet{Duan2025} proposed a storyline-analysis testbed for NeuralGCM and showed that it can reproduce the 2021 Pacific Northwest heatwave, while also noting limitations associated with the absence of an interactive land component. \citet{Baxter2026} assessed atmospheric variability in ACE2 and NeuralGCM and found that both models capture large-scale tropical-wave spectra but struggle with variability on quasi-biennial-oscillation timescales. More recently, \citet{Zhang_2025_NGCM} used NeuralGCM hindcasts to evaluate seasonal prediction of the tropical atmosphere and reported promising correlations with ERA5 and skill in year-to-year variability of monthly-mean fields.

Community testbeds have also emerged. ClimateBench \citep{Climatebench1} provides a standardized framework for comparing deep-learning Earth System emulators under long-range forcing scenarios, helping move evaluation beyond black-box performance. WeatherBench2 \citep{Rasp2024} provides standardized metrics and protocols for global short- to medium-range data-driven weather forecasting. These efforts offer important precedents, but they do not fully address the need for a comprehensive evaluation of DL-ESMs using the same diagnostics routinely applied to CMIP-class Earth System models.

Here, we address that gap by applying the PCMDI Metrics Package (PMP; \citep{lee2024PMP}) to a set of DL-ESMs. PMP provides a standardized and reproducible framework for evaluating ESM simulations against observational and reanalysis references, with diagnostics spanning mean climatology, modes of variability, precipitation variability, monsoon behavior, and extremes. Our goal is to compare selected DL-ESM outputs with reference observational datasets and with physics-based CMIP simulations that serve as benchmarks. In doing so, we build on the formal evaluation guidance of \citet{Ullrich2025} and demonstrate how PMP can help quantify the fit-for-purpose of DL-ESMs across variables, regions, and timescales.

This evaluation is also intended to inform broader emerging community efforts such as AIMIP (AI-based Model Intercomparison Project (\cite{AIMIP,henn2026aimip}) which aims to evaluate and standardize AI and Machine Learning approaches for Eart System modeling by comparing AI-based emulators through pre-defined experiments. By using established PMP diagnostics, our analysis provides a practical pathway for incorporating DL-ESMs into reproducible, and community-relevant model-evaluation workflows.

The paper is organized as follows. Section~\ref{Sect:ES} describes the selected models and evaluation strategy; Sect.~\ref{Sect:DP} summarizes the data preparation needed to compute PMP metrics applied in this study, outlines the boundary conditions, and discusses the need for data format standardization;
Section~\ref{Sect:PMP} presents the PMP diagnostics for the DL-ESMs and CMIP models. Finally, Sect.~\ref{Sect:disc} discusses the main findings and implications for future DL-ESM development.

\section{Selected models and evaluation strategy}
\label{Sect:ES}

Our study focuses on atmosphere-only DL-ESM configurations that can be run in an ``AMIP-style'' mode. In these simulations, the models are initialized from atmospheric states and then integrated forward using prescribed boundary conditions, allowing multi-year roll-outs that can be evaluated with Earth System-model diagnostics rather than only short-range forecast scores.

All analyzed DL-ESMs are forced with historical sea-surface temperature (SST) and sea-ice concentration (SIC) boundary conditions, following a strategy analogous to AMIP simulations with traditional physics-based models \citep{Gates1999}. We apply PCMDI PMP metrics to four DL-ESM configurations: one version of Ai2's ACE2 emulator \citep{wattmeyer2023} and three versions of Google's NeuralGCM \citep{Kochkov2024,Yuval2026}. These models differ both in architecture and in the way they represent atmospheric evolution.

ACE2 is a fully data-driven autoregressive emulator. Here we evaluate the ACE2 version trained with the ECMWF Reanalysis v5 dataset (ERA5; \citep{ERA5}) over 1940--2020; for simplicity, we refer to this configuration as ACE2 throughout the paper. ACE2 is forced by observed SST, sea ice, CO2, and solar radiation, which enables it to simulate historical atmospheric variability and forced long-term trends. Recent work \citep{watt2025ace2} has shown that this model can reproduce atmospheric warming trends, interannual variability in global-mean temperature, and the atmospheric response to El Niño SST variability. Because ACE2 is trained on historical reanalysis data, the present evaluation is not designed as a strict out-of-distribution extrapolation test. Instead, it asks whether a DL-ESM trained on historical data can meet the same observational and CMIP6-based performance benchmarks used for traditional models over a common historical period.

NeuralGCM follows a hybrid design that combines a differentiable physics-based dynamical core with learned parameterizations for unresolved processes, including processes related to convection and clouds. We evaluate three NeuralGCM configurations: (1) the original model introduced by \citet{Kochkov2024}, hereafter ``NeuralGCM''; (2) a later version described by \citet{Yuval2026}, hereafter ``NeuralGCM-evap'', that predicts precipitation minus evaporation to maintain consistency with the moisture budget but does not output precipitation directly; and (3) ``NeuralGCM-precip'', which is designed to output precipitation directly. The NeuralGCM-evap and NeuralGCM-precip configurations are trained with both ERA5 and Integrated Multi-satellitE Retrievals for the Global Precipitation Measurement (IMERG) precipitation data \citep{IMERG}, making them especially relevant for the precipitation-focused diagnostics discussed below.

Table~\ref{tab:models} summarizes the selected models and configurations. The original NeuralGCM includes a deterministic configuration available at $0.8^\circ$, $1.4^\circ$, and $2.8^\circ$ horizontal resolution, and a stochastic configuration at a $1.4^\circ$ resolution, in which random perturbations are introduced into the embedded physical parameterizations. NeuralGCM-precip and NeuralGCM-evap are inherently stochastic at a $2.8^\circ$ resolution, and do not have deterministic counterparts. In this study, we selected the three configurations of NeuralGCM based on their common  $2.8^\circ$ resolution.

%Differences between ACE and other DL-ESMs (specifically NeuralGCM).
\begin{table}[H]
    \centering
    \caption{Summary of the analyzed Deep-learning Earth System Models}
    \label{tab:models}
    % \begin{tabular}{>{\bfseries}l l}
    \begin{tabular}{llllll}
\tophline
         DL-ESM & Boundary Conditions & Reference & Training dataset & Mode & Resolution\\
\middlehline
                           ACE2 & \citep{Taylor2000_BC} & \citep{wattmeyer2023}& ERA5 & deterministic & $1^\circ$\\
         NeuralGCM & \citep{Taylor2000_BC} & \citep{Kochkov2024} & ERA5 & deterministic & $2.8^\circ$\\
                  NeuralGCM-evap & \citep{Taylor2000_BC} & \citep{Yuval2026} & ERA5, IMERG & stochastic & $2.8^\circ$ \\
                           NeuralGCM-precip & \citep{Taylor2000_BC} &  \citep{Yuval2026} & ERA5, IMERG & stochastic & $2.8^\circ$ \\
\bottomhline
\end{tabular}
\end{table}

% \section*{Stability and ensemble generation}
Our evaluation begins by generating single-model ensembles for each selected DL-ESM. ACE2 remained stable across the long roll-outs used here, so its ensemble could be produced by running the autoregressive model forward from varied initial conditions obtained from the official repository (\cite{ai2_2025} HuggingFace webpage), with start dates offset by one or more days. For the deterministic $2.8^\circ$ NeuralGCM configuration, which does not contain stochastic variability by default, we generated ensemble members by varying the initial conditions in a similar way, using the 15th day of each month from February through November 1979. For the stochastic NeuralGCM-evap and NeuralGCM-precip configurations, we additionally introduced perturbations through different random seeds.

Long-run stability is itself part of the evaluation. For NeuralGCM, we applied the correction provided by the model developers that keeps the global mean logarithm of surface pressure fixed. With this correction, we generated 10-member ensembles for ACE2 and the original NeuralGCM. The precipitation-capable NeuralGCM configurations were less robust. NeuralGCM-evap was generally more stable than NeuralGCM-precip, but the number of usable simulations depended on the selected end-date threshold. For example, using 31 December 2014 as the stability threshold yielded three stable NeuralGCM-evap runs and no stable NeuralGCM-precip runs. This difference in stability is important for interpreting the precipitation diagnostics, because some NeuralGCM-precip results are necessarily based on shorter or more limited samples.

We focus on key atmospheric output variables (listed in Table~\ref{tab:var}), using the ERA5 reference dataset \citep{ERA5}, for all the variables, except for precipitation, for which we use the listed Global Precipitation Climatology Project (GPCP) products \citep{Adler2018}.
%%%%%% Table 2  %%%%%%
\begin{table}[H]
\centering
\caption{List of output variables and reference observational datasets used for the evaluation.}
\small
\begin{tabular}{lllllll}
\tophline
Variable & Full name & Product & ACE2 & NeuralGCM & NeuralGCM-evap & NeuralGCM-precip \\
\middlehline
ta-200, ta-850 & Air temperature at 200 and 850 hPa & ERA5 & \checkmark & \checkmark & \checkmark & \checkmark  \\
ua-200, ua-850 & Zonal wind at 200 and 850 hPa & ERA5 & \checkmark & \checkmark & \checkmark & \checkmark  \\
va-200, va-850 & Meridional wind at 200 and 850 hPa & ERA5 & \checkmark & \checkmark & \checkmark & \checkmark \\
zg-200, zg-850 & Geopotential height at 200 and 850 hPa & ERA5 &  & \checkmark & \checkmark & \checkmark \\
pr & Precipitation, daily & GPCP 1.3 & \checkmark &  & \checkmark & \checkmark \\
pr & Precipitation, monthly & GPCP 2.3 & \checkmark&  & \checkmark & \checkmark \\
pr & Precipitation, daily$+$monthly & GPCP 3.2 & \checkmark&  & \checkmark & \checkmark \\
% ps & Air pressure at the surface & ERA5 & \checkmark & \checkmark & \checkmark & \checkmark  \\
psl & Sea level pressure & ERA5  & \checkmark* & \checkmark* &   &   \\

% psl & Sea level pressure & ERA5  & \checkmark\textdagger & \checkmark* &  ** & **  \\
% tas & 2 m air temperature & ERA5  & \checkmark & \checkmark & \checkmark & \checkmark \\
% tauu & Surface zonal wind stress & ERA-Interim & \citep{Dee2011} \\
% ts & Surface temperature & ERA5  & \checkmark & \checkmark & \checkmark & \checkmark \\
\bottomhline
\end{tabular}
\belowtable{* derived indirectly from surface pressure} % Table Footnotes
\label{tab:var}
\end{table}

We note that ACE2 and NeuralGCM do not directly provide sea level pressure (psl);
instead, this variable is calculated from each model's surface pressure. In ACE2, surface pressure is a prognostic variable in the core learning engine, whereas in NeuralGCM it is diagnostic and is not explicitly included in the loss function formulation.
NeuralGCM-evap  and NeuralGCM-precip currently do not provide surface pressure, and consequently do not provide sea level pressure.
% The evaluation of sea-level pressure (\textit{psl}) is complicated by the fact that this variable is not provided directly by the ACE2 and NeuralGCM outputs. Estimating psl through standard vertical interpolation or extrapolation can introduce additional uncertainty, particularly over regions of complex topography where the hydrostatic correction is sensitive to temperature structure and surface elevation.
We found that the simulation of psl has been challenging and so has been deriving this variable indirectly via interpolation as it is not immediately an output in ACE2 and NeuralGCM models.
To reduce possible errors introduced by the specific interpolation method, we developed an \textit{ad-hoc} machine learning strategy to infer monthly mean sea-level pressure for models that do not provide it directly.
Since the hydrostatic relationship is non-linear, a residual convolutional neural network was used to predict monthly mean sea-level pressure from gridded monthly mean atmospheric features. The input predictors included surface pressure and near-surface temperature, together with static geographic information such as elevation, latitude, and longitude, to help represent spatial variations associated with topography and location.
Preliminary sea-level pressure results (included in Figure~\ref{fig:portrait_RSME} in Section \ref{Sect:PMP}) are encouraging, suggesting this approach may offer a practical pathway for including missing sea-level pressure diagnostics in future evaluations of DL-ESMs.
% % \textdagger Sea level pressure is not directly available for ACE2 and needs to be calculated offline from surface pressure, which is a prognostic variable 
% ** Surface pressure is not currently available from either NeuralGCM-evap or NeuralGCM-precip, preventing the computation of sea-level pressure. \\
\section{Data Preparation}
\label{Sect:DP}
The PCMDI Metrics Package (PMP) is an open-source Python framework designed to evaluate the consistency between ESMs and observational data using established statistical methods \citep{lee2024PMP}. PMP is integrated with the CMIP evaluation framework and has been used primarily to track the performance of physics-based Earth System models across CMIP generations. It can be applied to fully coupled CMIP historical simulations and to AMIP simulations, in which the atmosphere is run with prescribed observed SST and SIC boundary conditions.

Because the DL-ESMs evaluated here are atmosphere-only systems, we apply the AMIP-oriented PMP diagnostics and use the corresponding observational products as reference datasets. These products are processed to be compatible with Observations for Model Intercomparison Projects conventions \citep{Obs4MIPs}. Although the selected DL-ESMs were originally trained with ERA5, the simulations analyzed here use SST and SIC boundary conditions from the PCMDI-AMIP forcing dataset \citep{Gates1999}. This choice is essential for a fair comparison with CMIP6 AMIP simulations, which are forced with the same lower-boundary-condition dataset. It also introduces a modest generalization test, because the boundary-condition source differs from the ERA5 training environment.

For ACE2, additional required forcings, including land fraction and surface geopotential height, were obtained from the official documentation \citep{wattmeyer2023}. ACE2 and NeuralGCM both advance at 6-hourly time steps. However, prescribing 6-hourly SST and SIC would require assumptions about the sub-daily evolution of fields that are normally supplied at coarser temporal resolution. We therefore reprocessed monthly SST and SIC following \citet{Taylor2000_BC} and interpolated them to daily values while preserving the original monthly means.
To provide a consistent benchmark against traditional models, we recomputed selected PMP metrics for a subset of CMIP6 simulations. We used one realization for climatology diagnostics and multiple realizations where needed for variability diagnostics for both CMIP-type and DL-ESMs. Model and reference fields were evaluated over maximally overlapping historical periods. Because PMP requires standardized model output, we converted the DL-ESM outputs into a CMIP-like structure using the same conventions applied to physics-based models, including Climate Model Output Rewriter (CMOR; \cite{CMOR3})-style variable names, metadata, and dimensions where possible.

\section{Applying PMP Metrics and Diagnostic Tools to DL-ESMs}
\label{Sect:PMP}
The PMP framework computes hundreds of summary statistics, including mean-state metrics, variability metrics, and process-oriented diagnostics.
In the context of ESM evaluation, the PMP toolkit has been critical for quantifying model performance against observed climatological fields and major modes of variability.
Adopting PMP for DL-ESMs allows us to evaluate these models with the same diagnostics used for CMIP-class simulations, while also exposing model-specific limitations such as missing output variables, post-processing requirements, and stability constraints.
We therefore organize the analysis into diagnostic categories that test complementary aspects of Earth system performance: mean climatology, modes of extra-tropical and intra-seasonal variability, monsoon behavior, and precipitation variability.

\subsection{Climatology Metrics}
A first way to assess the fitness-for-purpose of the DL-ESMs is to examine how well they reproduce the large-scale mean Earth system state. We therefore compute annual and seasonal area-weighted climatologies for the key variables listed in Table~\ref{tab:var}.

Among the variables considered here, precipitation is one of the most challenging to simulate in both traditional ESMs and DL-ESMs because it depends on interactions among circulation, moisture, clouds, convection, and surface fluxes across a wide range of scales \citep{Stephens2010,Sonderby2020MetNet}.

%---------------------- FIGURE 1 ------------------------------------------
%% ACE bias
\begin{figure}[H]
    \centering
    \includegraphics[width=0.9\textwidth]{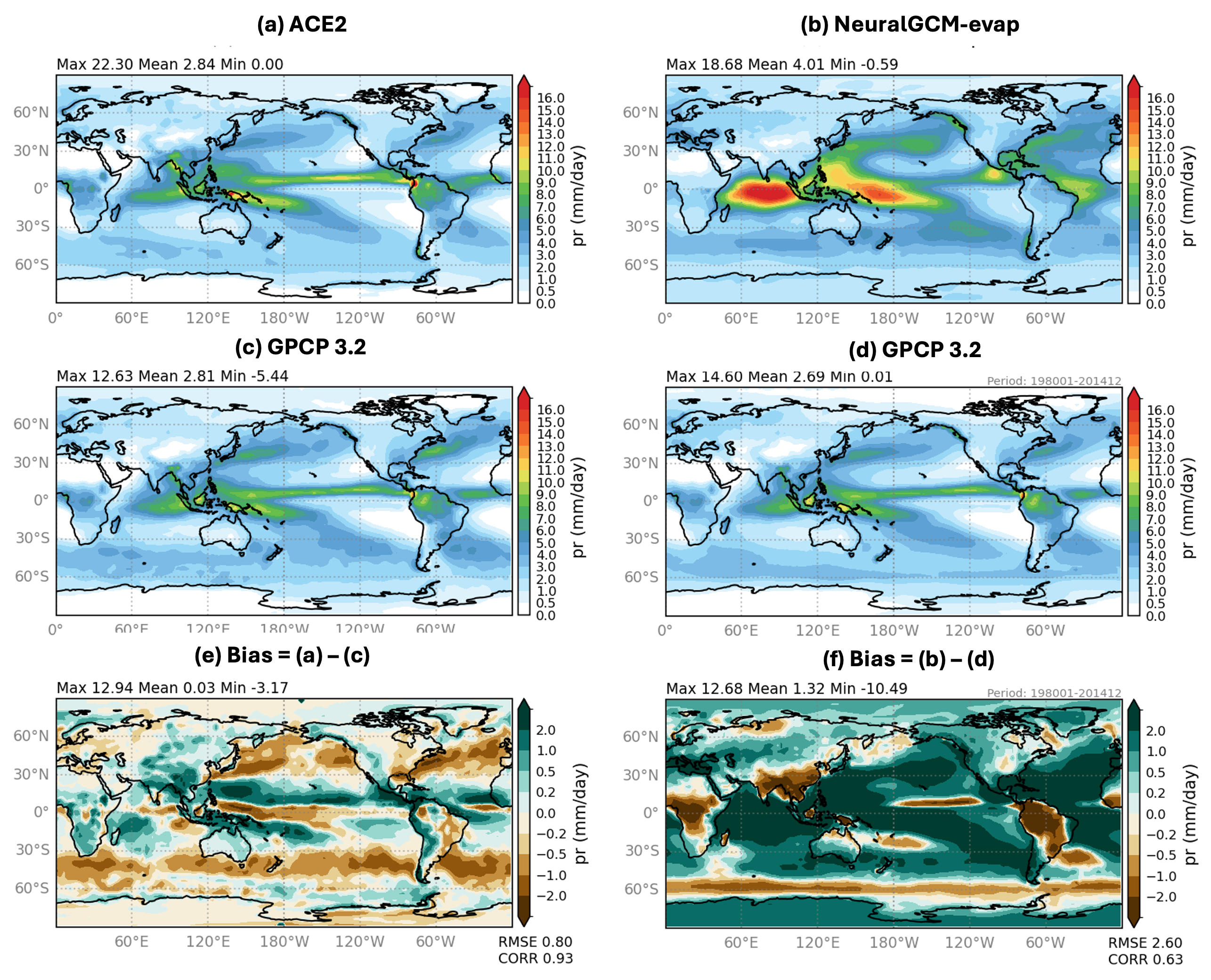}
    \caption{Annual-mean precipitation climatology and bias for ACE2 and NeuralGCM-evap. Panels (a) and (b) show model climatologies; panels (c) and (d) show the corresponding GPCP 3.2 reference climatologies \citep{Adler2018}; and panels (e) and (f) show model-minus-GPCP biases. Statistics in the bias panels summarize RMSE and spatial correlation against GPCP for the analysis periods are shown in the panels.}
    \label{fig:pr_bias}
\end{figure}
%--------------------------------------------------------------------------

Figure~\ref{fig:pr_bias} illustrates an important first test of whether the DL-ESMs reproduce the observed annual-mean precipitation distribution.
The figure compares ACE2 and NeuralGCM-evap with GPCP 3.2 and shows both the simulated climatologies and the model-minus-observation biases.
ACE2 reproduces the broad observed distribution of tropical rainfall, including the main rain belts over the equatorial Pacific, equatorial Atlantic, and South Pacific Convergence Zone.
However, its bias field (Fig.~\ref{fig:pr_bias}e) shows a coherent wet bias along the tropical convergence zones and adjacent high-evaporation oceanic regions.
This pattern suggests that ACE2 does not simply displace the main tropical rain bands; rather, it tends to over-intensify or broaden climatological convergence-zone precipitation.
The high spatial correlation shown in the figure (0.93) indicates that the large-scale pattern is realistic, while the remaining RMSE (0.80~mm~day$^{-1}$) highlights persistent tropical amplitude errors.
This interpretation is consistent with recent analyses of ACE2 trained on ERA5, which also identify the largest time-mean precipitation errors in the ITCZ and South Pacific Convergence Zone regions \citep{watt2025ace2}.
It is also consistent with broader benchmarking work showing that AI emulators can struggle to represent longer-timescale processes and feedbacks beyond those directly emphasized by the training loss \citep{Baxter2026}.

The NeuralGCM-evap precipitation climatology shows larger errors.
Because this model version is trained to predict precipitation minus evaporation rather than precipitation as an independent output, precipitation must be inferred indirectly from the moisture budget.
The resulting bias field (Fig.~\ref{fig:pr_bias}f) shows a widespread tropical wet bias and substantially weaker agreement with GPCP than ACE2, with a higher RMSE (2.60~mm~day$^{-1}$) and lower spatial correlation (0.63).
Part of this degradation may reflect the shorter stable analysis period available for NeuralGCM-evap and sensitivity to the prescribed SST sequence, but the result also highlights precipitation as a key target for DL-ESM development.

% Metric visualization is central to diagnosing model performance. To visualize the performance of the model, several diagnostic tools can be used as in \citep{lee2024PMP}.
% It is insightful to visualize the mean climate metrics via multiple plots to facilitate the comparison between model fields, sharing the same observational reference.
A complete set of annual and seasonal climatology fields for NeuralGCM and ACE2, for all variables reported in Table~\ref{tab:var}, is provided in the Supplement (Figs.~\ref{SIfig:clim_ACE_ta200}--\ref{SIfig:clim_ACE_pr}).

To efficiently summarize climatology metrics in a compact and informative way, PMP 
uses portrait plots to display normalized global spatial Root Mean Square Error (RMSE) of seasonal climatologies of various meteorological variables for each tested model \citep{gleckler2008performance}. The normalization allows for an easier comparison across variables having different absolute bias magnitudes across models.
%---------------------- FIGURE 2 ------------------------------------------
\begin{figure}[H]
    \centering
    \includegraphics[width=1\textwidth]{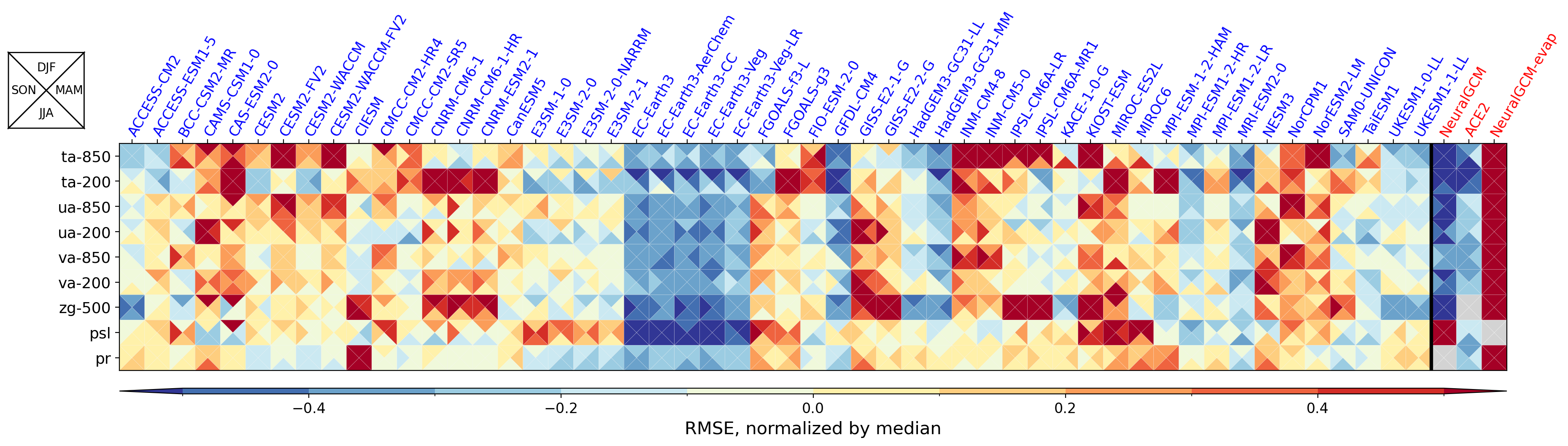}
    \caption{Portrait plot for normalized spatial Root Mean Square Error (RMSE) across different seasons. Negative normalized error indicates performance better than the multi-model median (like in \citep{Ullrich2025} and \citep{lee2024PMP}). The climatology metric is computed with respect to the observed climatological fields provided by the reference dataset product corresponding to each examined variable as reported in Table~\ref{tab:var}. The RMSE is calculated for each season (shown as triangles in each cell) over the globe, for the period 1981-2013. Both model and reference data were interpolated to a common $2.5^\circ \times 2.5^\circ$ grid.
    Grey cells correspond to unavailable output variables due to the specific model design.  
    The RMSE of each variable is normalized by the median RMSE of all CMIP6 models and DL-ESMs considered. For instance, a normalized RMSE of 0.4 corresponds to an error 40\% greater than the median RMSE across all models.}
    \label{fig:portrait_RSME}
\end{figure}
%--------------------------------------------------------------------------
Figure~\ref{fig:portrait_RSME} shows the normalized RMSE for seasonal averages of the variables listed in Table~\ref{tab:var} for all the tested CMIP6 models and DL-ESMs.
By using a common color scale for all variable statistics and models, we can see how the DL-ESMs compare to the CMIP6 in terms of their normalized RMSE using the corresponding reference dataset per variable, thus highlighting the relative strengths of different models.
We can also compare across DL-ESMs, noting that NeuralGCM-evap produces consistently larger normalized errors across variables and seasons, whereas the original NeuralGCM exhibits variable-dependent performance, with normalized RMSE values generally below the multi-model median. 
This earlier model version does not have the capability to output precipitation (which is why is shown as a missing metric in Figure~\ref{fig:portrait_RSME}).
However, NeuralGCM is showing better performance compared to NeuralGCM-evap because it is trained on a relatively longer time period, making NeuralGCM-evap's results more challenging as they are derived via extrapolation, which can partially explain the obtained performance.
Moreover, NeuralGCM-precip is not included in the portrait plot results as the model showed lack of stability for the selected climatology period. Its performance is included in the subsequent analysis where the computation of the metrics (e.g., monsoon) did not cover the whole time period.
We also notice the good performance of ACE2 across all the output variables.
Fig.~\ref{SIfig:portrait_MAE}, which displays the portrait plot for the Mean Absolute Error instead of the normalized spatial Root Mean Square Error, shows similar results.
% It is noteworthy to recall that ACE2 output variables are provided on model levels rather than standard pressure levels, and we interpolated them to pressure levels in log-pressure space using the nine available ACE vertical levels. We note, however, that the results were sensitive to the choice of interpolation method, and the associated uncertainty may degrade model skill at 850 hPa, especially for ta850. In order to improve skill, we adopted a masking approach for the times and locations where the model surface pressure was lower than the requested pressure level and recomputed the climatology. This strategy allowed to reduce the vertical interpolation errors we initially encountered. As an illustrative example, the resulting improved bias map for t850 is reported in Fig.~\ref{SIfig:ACEbias_ta850}.
It is important to note that ACE2 atmospheric output is provided on native model levels rather than on standard pressure levels. For comparison with observational products and CMIP-style diagnostics, we interpolated these fields to pressure levels in log-pressure space using the nine available ACE2 vertical levels. We found that the resulting fields were sensitive to the interpolation method, introducing an additional source of uncertainty that can degrade apparent model skill at 850~hPa, especially for ta-850. To mitigate this issue, we applied a surface-pressure mask prior to computing the climatology, excluding all times and locations for which the model surface pressure was lower than the requested pressure level. This procedure removes below-surface values and substantially reduces the vertical interpolation artifacts identified in the initial analysis. An illustrative example of the improved ta-850 bias field is shown in Fig.~\ref{SIfig:ACEbias_ta850}.
% Parallel coordinate plots allow to compare multiple diagnostic metrics at once to identify systematic biases or compensating errors.

\begin{figure}[H]
    \centering
    \includegraphics[width=0.9\textwidth]{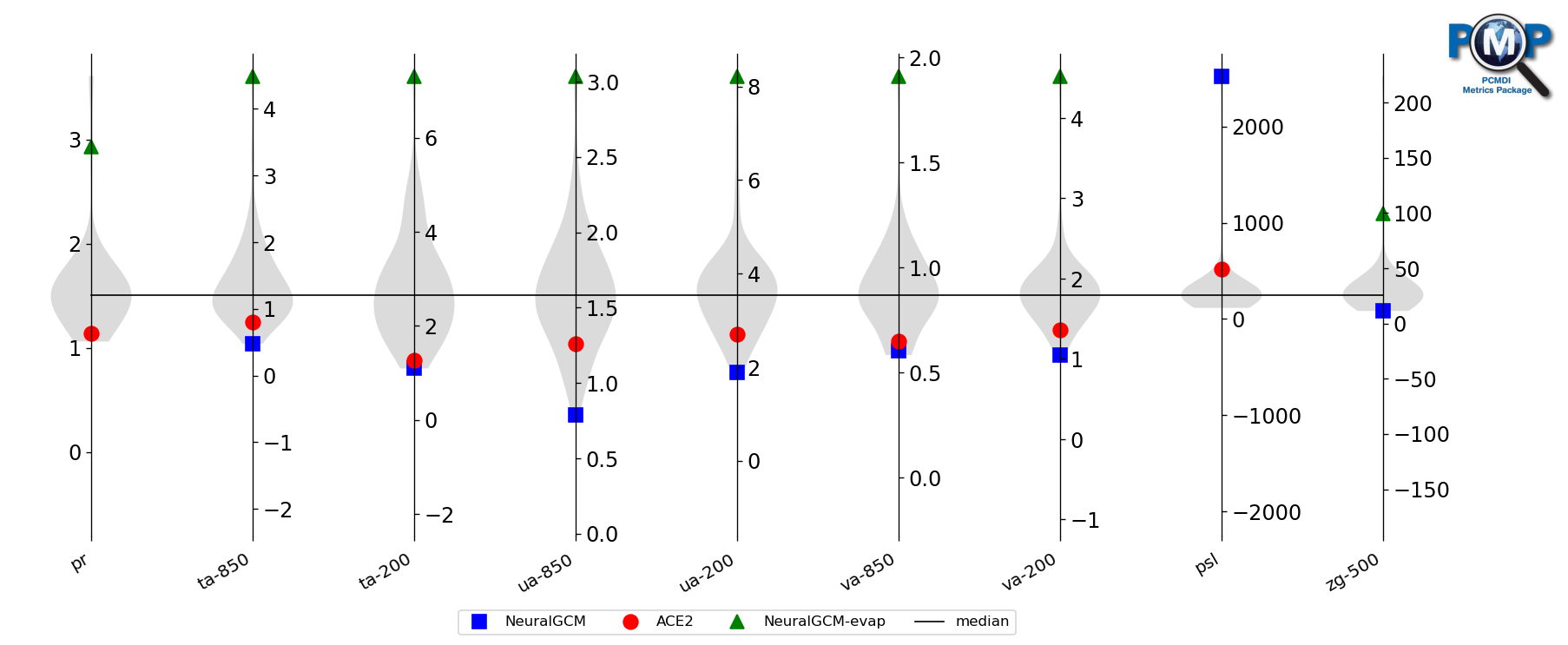}
    \caption{Parallel coordinate plot for spatiotemporal RMSE from PMP mean climatology.
Each vertical axis represents a different variable. The distributions of RMSE from CMIP6 modes are visualized as violin plots shaded in gray. The colored markers represent the DL-ESMs: NeuralGCM, ACE2, and NeuralGCM-evap. The time epoch used for this analysis is 1981–2013.
The middle of each vertical axis is aligned with the median statistic of all CMIP6 models and DL-ESMs. }
\label{fig:parallel_RSME_clim}
\end{figure}

To complement the portrait plot of relative performance, we also use a parallel coordinate plot for spatiotemporal RMSE, as in \citep{lee2024PMP}, to show the absolute error value (Figure~\ref{fig:parallel_RSME_clim}). This plot reveals that ACE2 and NeuralGCM provide errors comparable to the CMIP6 models for most of the analyzed variables, consistent with the previously presented results. In contrast, NeuralGCM-evap shows larger spatiotemporal errors in climatology for most variables compared to CMIP6 models. These errors could be due to the relatively shorter training period, given the stability issues for the model on longer time periods.
%---------------------- FIGURE 3 --------------------------
%--------------------------------------------------------------------------

\subsection{Extra-tropical Modes of Variability Diagnostics}

The PMP framework also provides diagnostics for evaluating model performance in representing major modes of climate variability. We omit a detailed assessment of the El Niño--Southern Oscillation (ENSO) in the present analysis because sea surface temperatures are prescribed in these simulations, so ENSO variability is imposed through the boundary conditions rather than freely generated by the models. This expectation was confirmed by comparing the simulated Niño 3.4 index with the reference index: both ACE2 and NeuralGCM reproduce Niño 3.4 variability very closely, with correlations greater than 0.95.

We therefore focus our variability assessment on diagnostics that more directly test the Extra-tropical modes of variability (ETMoV), both in terms of pattern and amplitude, using the common basis function (CBF) approach introduced in \citep{Lee2019a}.
More specifically, among the the PMP's ETMoV metrics, we focus on four sea-level-pressure-based modes: the Southern Annular Mode (SAM), the Northern Annular Mode (NAM), the North Atlantic Oscillation (NAO) and the Pacific North America pattern (PNA). 

\begin{figure}[H]
    \centering
    \includegraphics[width=1\textwidth]{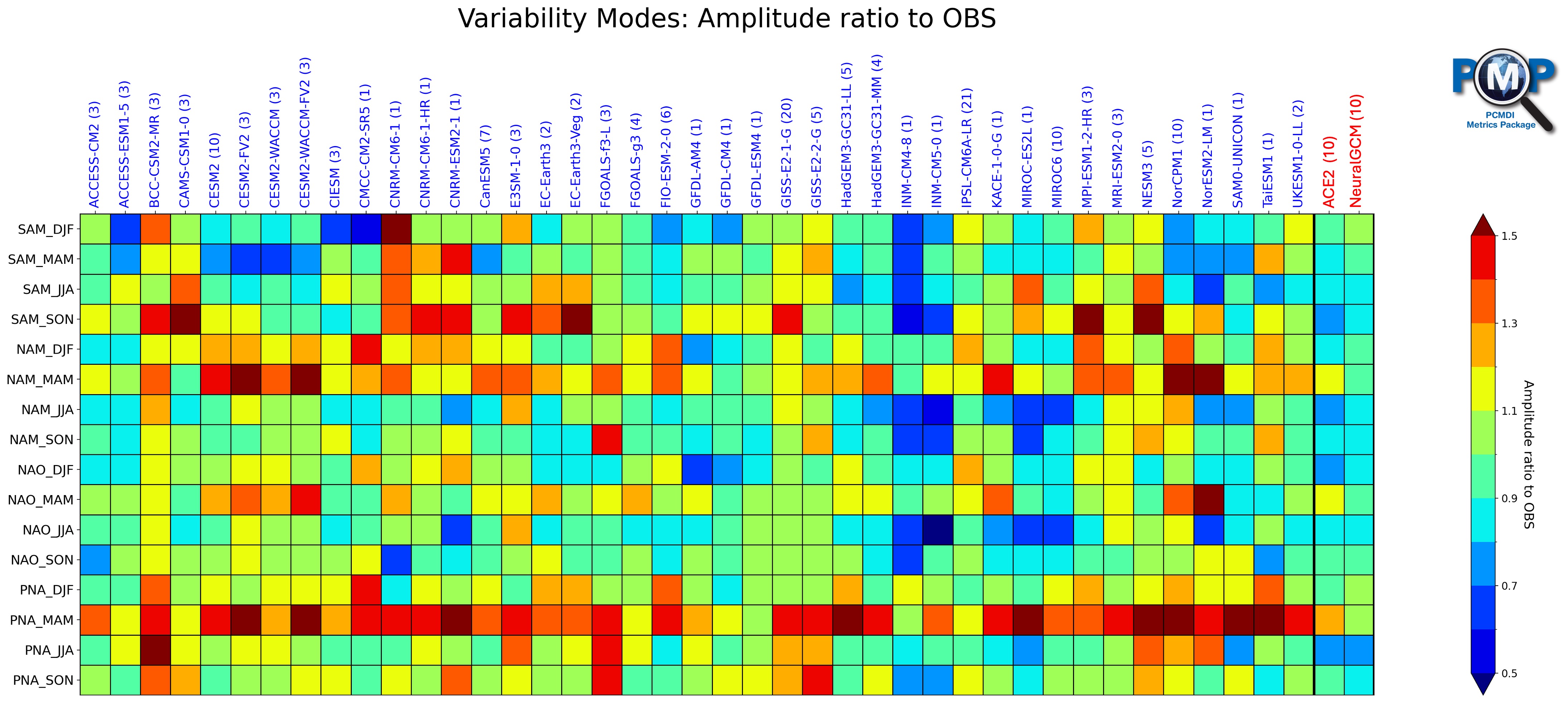}
    \caption{Portrait plot of the amplitude of extra-tropical modes of variability simulated by CMIP6 models and DL-ESMs in their historical or equivalent simulations.
    The amplitude ratio metric is the ratio of spatiotemporal standard deviations of the model versus the observed principal components (PCs), obtained using the CBF method in the PMP \citep{lee2024PMP}. Rows correspond to mode and season, and columns to models, being the last two columns for the DL-ESMs, separated by a vertical thick black line. Results are for four sea-level-pressure-based modes (SAM, NAM, NAO, and PNA) and are relative to NOAA-20CR. The number of ensemble members per model is reported in parentheses after the model name. Metric value per model is obtained by averaging the metrics for individual ensemble members.
    % \textbf{Two time periods}: 1980--2005 for the current model and 1900--2005 from 20CR (for historical consistency).
    }
    \label{fig:portrait_MoV_ampl}
\end{figure}

\begin{figure}[H]
    \centering
    \includegraphics[width=0.9\textwidth]{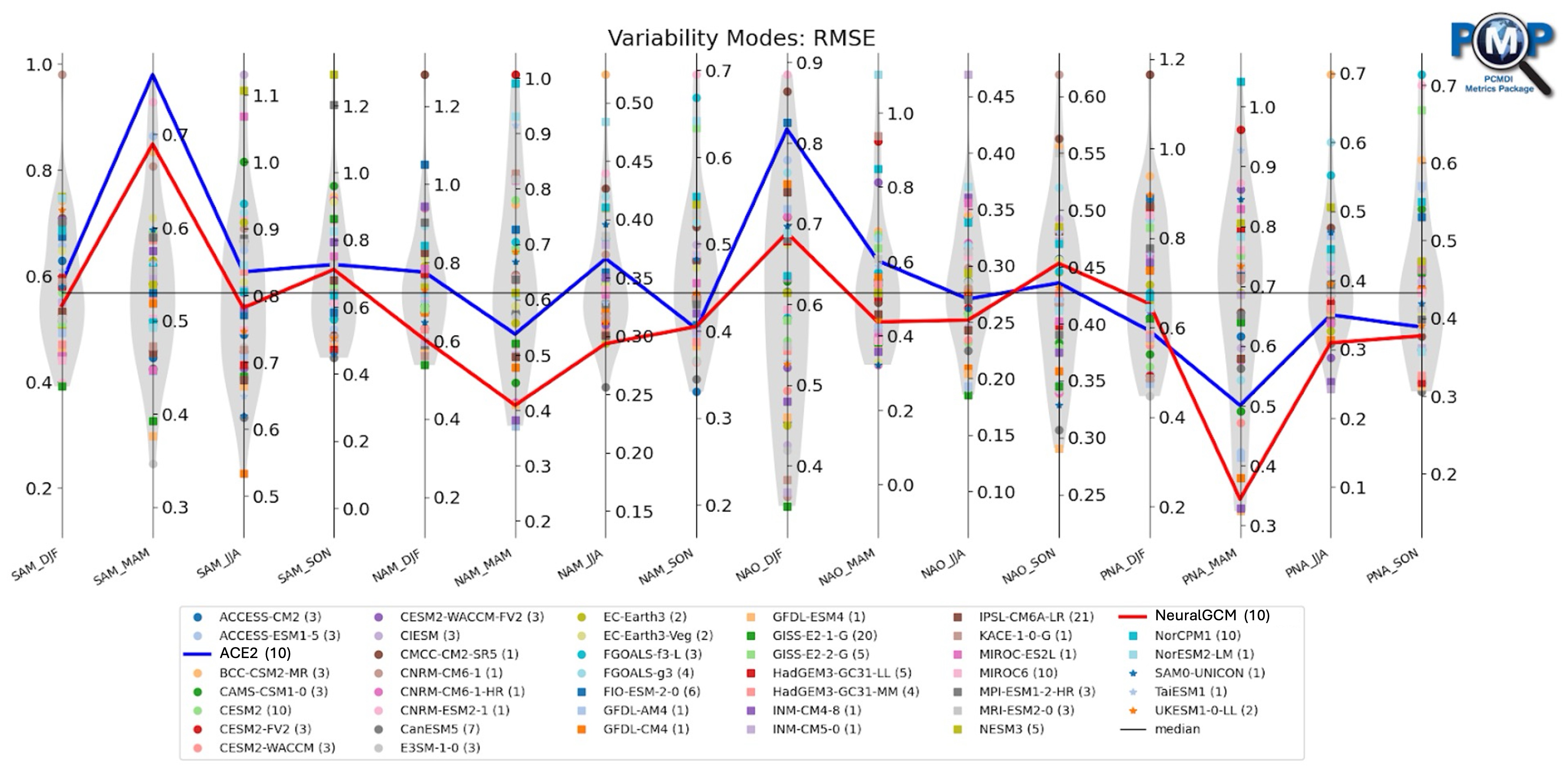}
    \caption{Parallel coordinate plot for spatiotemporal RMSE for CMIP6 model ensembles (gray shaded) with individual models plotted as colored markers. The metrics for the two DL-ESMs are shown as solid lines (ACE2 in blue and NeuralGCM in red).}
    \label{fig:parallel_RSME_MoV}
\end{figure}

\begin{figure}[H]
    \centering
    \includegraphics[width=0.9\textwidth]{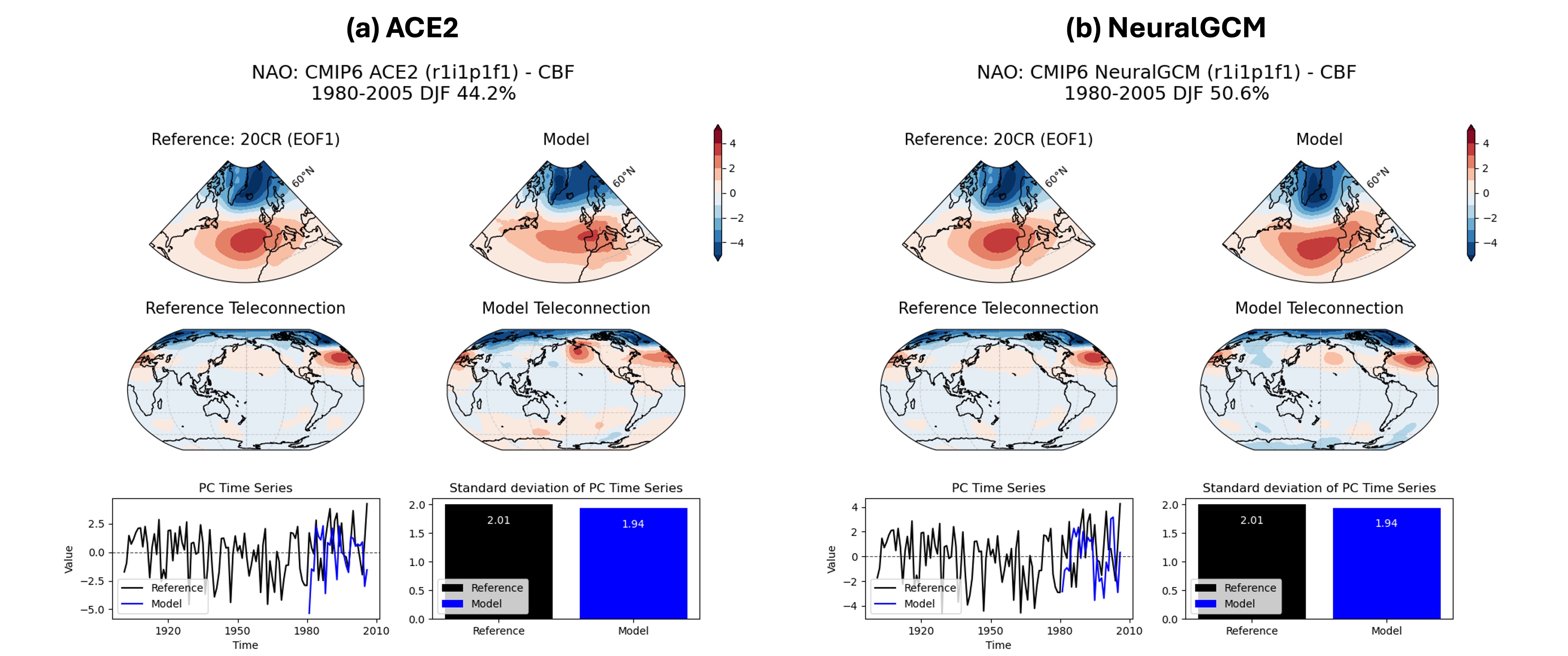}
    \caption{NAO skill comparison in DJF period: (a) ACE2 simulation as compared to the observations in the NAO domain (top row), ability to reproduce the associated teleconnections (middle row) and time series comparison, with a representation of the standard deviations from the model and the observations (bottom row).}
    \label{fig:MoV_NAO}
\end{figure}

In Figure~\ref{fig:portrait_MoV_ampl} we report the amplitude metric, defined as the ratio between the standard deviations of the model and observed principal components, per single-model ensemble \citep{Lee2019a}.
Metric values close to unity are a measure of good agreement with the reference observational dataset.
We notice that green shading predominates in the table, mostly in CMIP6 models and also in the two tested DL-ESMs, indicating similar variances to observations.
Among the DL-ESMs, NeuralGCM exhibits more uniform performance across the ETMoV and seasons whereas ACE2 shows more variability across the ETMoV. In particular, ACE2 has slightly larger variance than observations for NAM Spring averages and lower variance in SAM Fall, NAM Summer and NAO Winter averages, while still showing overall satisfactory performance.
A very recent paper \citep{Baxter2026} provides insights about the ability of ACE2 and NeuralGCM to reproduce complex atmospheric dynamics, including eddy-mean flow interactions which are critical for extra-tropical modes of variability. The study concludes that both DL-ESMs are able to simulate extra-tropical wave-mean flow interactions but they struggle with the propagation of the
SAM. In another recent paper \citep{Kent2025} the authors show that ACE2 model exhibits skillful seasonal predictions of NAO.
In other ETMoV cases, such as for PNA Spring, most CMIP6 models overestimate the observed amplitude along with ACE2, while NeuralGCM seems to capture better the observed amplitude.
%---------------------- FIGURE 4 --------------------------
%--------------------------------------------------------------------------
Figure~\ref{fig:parallel_RSME_MoV}, showing the spatiotemporal RMSE in the format of parallel coordinate plot, indicates that both ACE2 and NeuralGCM  perform consistently with the CMIP6 models in simulating ETMoV, with particularly consistent skill for NAO across seasons.
% more insights from literature
%---------------------- FIGURE 5 --------------------------
%--------------------------------------------------------------------------
Figure~\ref{fig:MoV_NAO} shows the December--January--February (DJF) NAO pattern simulated by ACE2 and NeuralGCM over the North Atlantic domain, confirming that both DL-ESMs capture this mode well (top row). NeuralGCM shows slightly superior ability to reproduce NAO in the Winter period. This could be due to the fact that NeuralGCM hybrid architecture simulates the NAO through a framework grounded in established physical laws, supplemented by AI, while ACE2 learns to emulate the patterns and variability of the NAO purely from data.

%---------------------- FIGURE 6 --------------------------
%--------------------------------------------------------------------------
\subsection{Intra-seasonal Oscillation Metrics}
\label{Sect:MJO}

The PMP framework offers several tools to study how models reproduce important features of the Earth system, such as the Madden--Julian Oscillation (MJO).
MJO is the most prominent intra-seasonal mode of variability in the tropics and its presence can affect tropical cyclone formation and lead to variations in rainfall and surface temperatures. The MJO is characterized by a slow eastward phase speed, a planetary zonal scale and a period of 30–60 days. 

Following \citet{lee2024PMP}, we compute the east--west power ratio (EWR) and eastward power normalized by observations (EOR) from wavenumber--frequency spectra of equatorial precipitation.
EWR is the ratio of spectral power in the eastward-propagating MJO band (zonal wavenumbers 1--3, periods of 30--60 days) to the corresponding westward-propagating power.
Values greater than one indicate stronger eastward than westward propagation, while values closer to the observational reference indicate a more realistic propagation amplitude.
EOR measures the eastward power normalized by the observed GPCP v1.3 power over the same MJO band, following \citet{Ahn2017}.

\begin{figure}[H]
    \centering
    \includegraphics[width=1\textwidth]{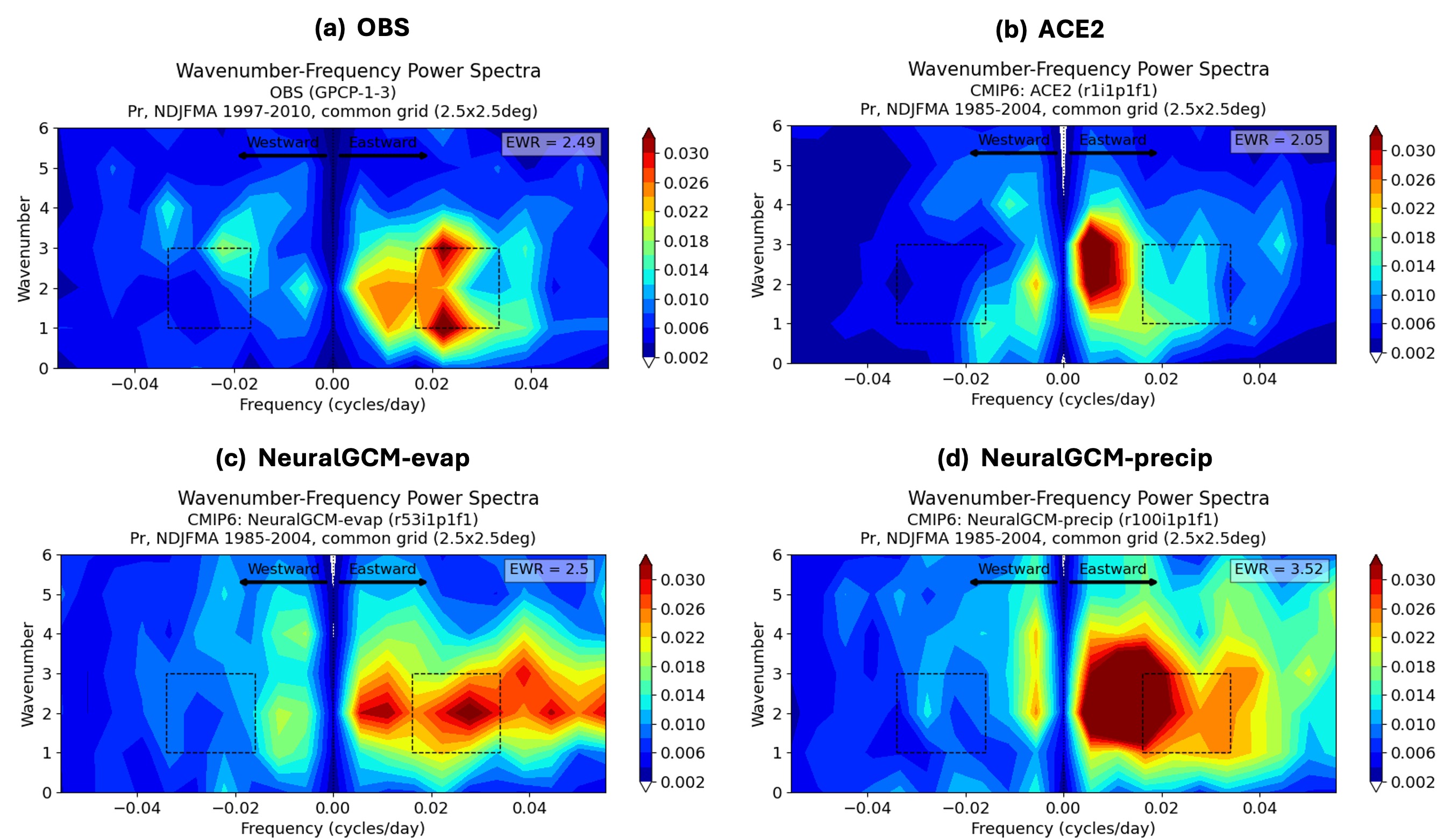}
    \caption{MJO propagation diagnostics for boreal winter (November--April). Wavenumber--frequency power spectra of daily precipitation averaged over $10^\circ$S--$10^\circ$N are shown for (a) GPCP 1.3 observations \citep{Huffman2001}, (b) ACE2, (c) NeuralGCM-evap, and (d) NeuralGCM-precip. Dashed boxes mark the westward- and eastward-propagating MJO bands used to compute the EWR. Spectra are averaged over all available years for each dataset; power units are mm$^2$~day$^{-2}$ per frequency interval per wavenumber.}
    \label{fig:MJOprop_winter}
\end{figure}

Figure~\ref{fig:MJOprop_winter} shows boreal-winter (November--April) wavenumber--frequency spectra for GPCP observations and the three DL-ESMs that provide precipitation.
All three DL-ESMs show an eastward-propagating MJO signal, but the balance between eastward and westward power differs by model.
NeuralGCM-evap most closely matches the observed EWR (2.49), with an EWR of about 2.5.
ACE2 has a weaker but still eastward-dominant signal (EWR~=~2.05), whereas NeuralGCM-precip produces a stronger east--west contrast (EWR~=~3.52).
Thus, the precipitation spectra indicate that the DL-ESMs can represent the directionality of MJO propagation, but differ in the amplitude and spectral distribution of that propagation.
The May--October results are shown in Fig.~\ref{SIfig:MJOprop_summer}; for that season, ACE2 (EWR~=~2.76) is closest to the observed EWR of 3.05.

%---------------------- FIGURE 7 --------------------------
%---------------------------------------------------------

\begin{figure}[H]
    \centering
    \includegraphics[width=1\textwidth]{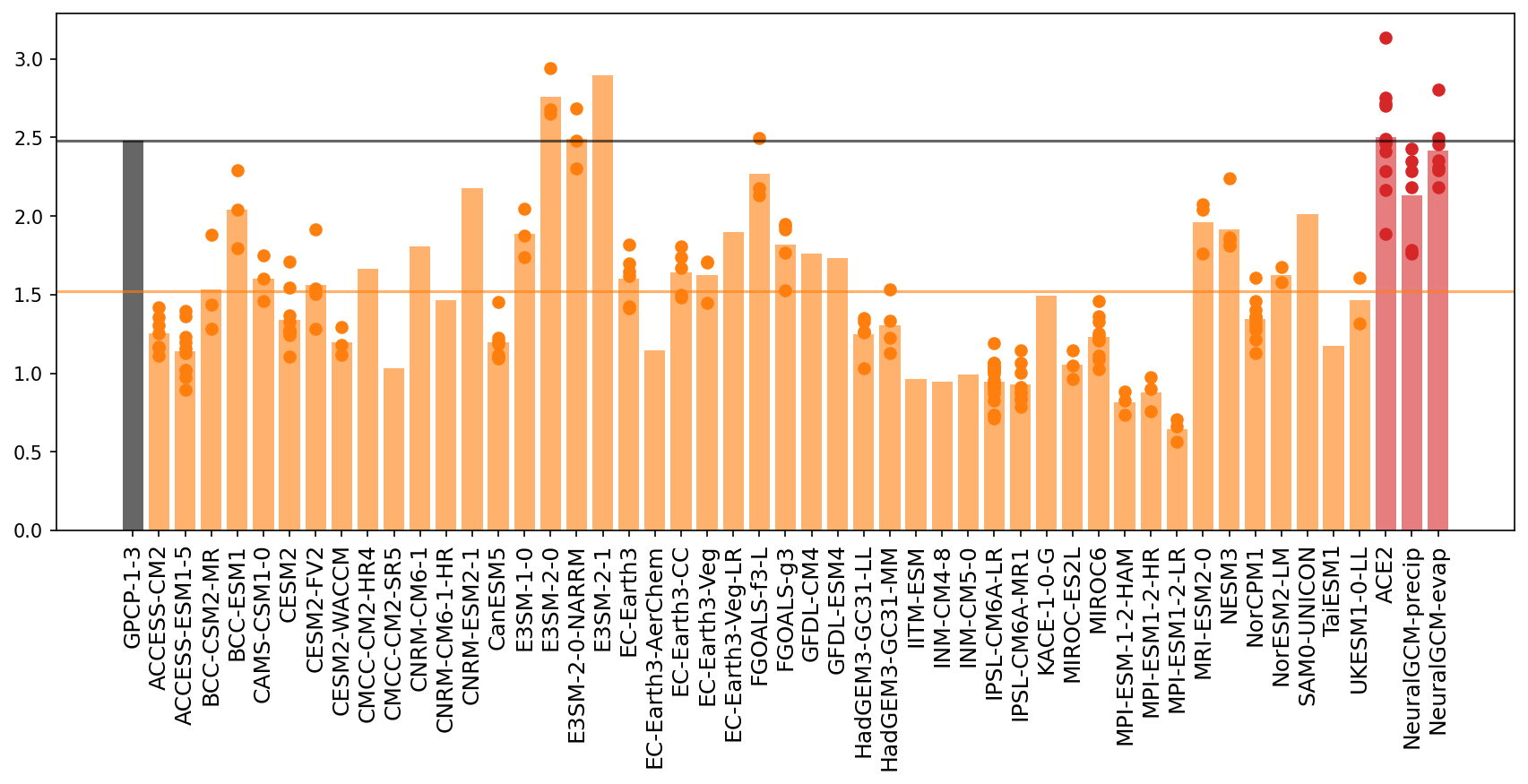}
    \caption{MJO east--west power ratio (EWR; unitless) for boreal winter. Orange bars show CMIP6 model means, red bars show DL-ESM means, and dots indicate individual ensemble members where available. The gray bar and black horizontal line denote the GPCP 1.3 observational reference \citep{Huffman2001}. Ensemble members and sample sizes follow the convention used in Fig.~\ref{fig:portrait_MoV_ampl}.}
    \label{fig:MJO_ensembles_winter}
\end{figure}

Figure~\ref{fig:MJO_ensembles_winter} places the DL-ESM EWR values in the context of CMIP6 single-model ensembles.
The CMIP6 models show large inter-model and intra-model spread, consistent with previous analyses of MJO metrics \citep{Back2024}, with only a subset of models (e.g., E3SM and FGOALS models) lying close to the GPCP reference.
Such spread reflects the known difficulty of simulating realistic MJO propagation in traditional models; previous work has linked MJO propagation errors to biases in the lower-tropospheric mean moisture gradient and zonal winds, as well as to tropical rain-belt biases \citep{Xiang2017,Jiang2020}.
Against these physics-based benchmarks, ACE2 and NeuralGCM-evap show competitive winter EWR values, while NeuralGCM-precip tends to overemphasize eastward relative to westward power.
Our results corroborate recent studies reporting realistic MJO behavior in ACE2 \citep{Chien_2025,watt2025ace2} and skillful S2S-scale predictions of MJO propagation and North Pacific circulation in NeuralGCM \citep{Peings_2026_NGCMS2S}.

%---------------------- FIGURE 8 --------------------------

The May--October EWR ensemble summary is shown in Fig.~\ref{SIfig:MJO_ensembles_summer}. It shows broadly similar behavior for CMIP6 models and DL-ESMs, with ACE2 and NeuralGCM-evap remaining skillful relative to the GPCP reference. These comparisons can be potentially affected by the ensemble-size differences, because the DL-ESM ensembles contain 10 members, whereas several CMIP6 models contribute to the analysis with fewer members.

\subsection{Regional Process Diagnostics: Monsoon skill}  
Monsoon diagnostics provide a regional and process-oriented test of precipitation timing, amplitude, and spatial organization.
Many CMIP models struggle to represent monsoon rainfall accurately, and the same diagnostics are useful for identifying whether DL-ESMs reproduce the seasonal accumulation and geographic extent of monsoon precipitation.
PMP includes metrics for monsoon onset, decay, and duration based on fractional accumulated precipitation, following \citet{Sperber2004}.
These metrics are useful because they diagnose phase errors and timing biases rather than only seasonal-mean rainfall totals.

\begin{figure}[H]
    \centering
    \includegraphics[width=0.8\textwidth]{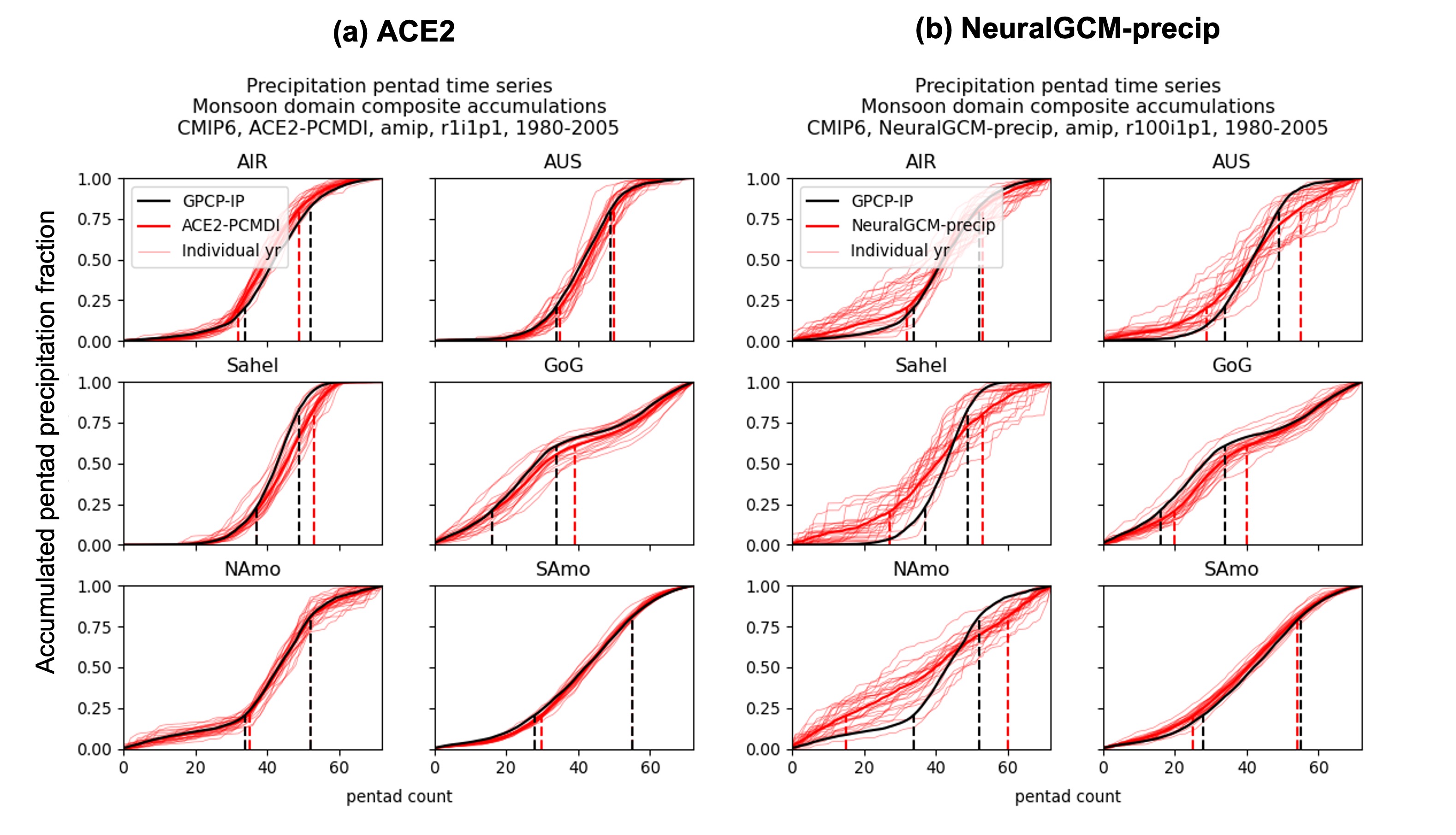}
    \caption{Pentad precipitation accumulation for regional monsoon diagnostics. Curves show fractional accumulated precipitation in observations and model simulations for six monsoon regions: all-India rainfall (AIR), northern Australia (AUS), Sahel, Gulf of Guinea (GoG), North American monsoon (NAMo), and South American monsoon (SAMo), for 1980--2005. Observational reference is GPCP 1.3; simulations are from (a) ACE2 and (b) NeuralGCM-precip. Vertical dashed lines mark observed (black) and modeled (red) onset and decay dates.}
    \label{fig:monsoon_Sper}
\end{figure}

As in \citet{lee2024PMP}, we compute area-averaged precipitation for six monsoon regions: all-India rainfall (AIR), northern Australia (AUS), Sahel, Gulf of Guinea (GoG), North American monsoon (NAMo), and South American monsoon (SAMo).
Figure~\ref{fig:monsoon_Sper} compares the accumulated pentad precipitation fractions for ACE2 and NeuralGCM-precip against observations.
We find that ACE2 generally follows the observed accumulation curves and shows a narrower spread across ensemble members, indicating more stable monsoon timing from year to year.
In contrast, NeuralGCM-precip shows larger year-to-year spread in most regions and tends to produce an earlier monsoon onset, except in the Gulf of Guinea region.
This early onset could reflect either an overly strong annual cycle or a phase advance in the simulated seasonal cycle.
NeuralGCM-evap produces similar behavior and is shown in Fig.~\ref{SIfig:monsoon_Sper_evap}.

%---------------------- FIGURE 9 --------------------------

\begin{figure}[H]
    \centering
    \includegraphics[width=0.7\textwidth]{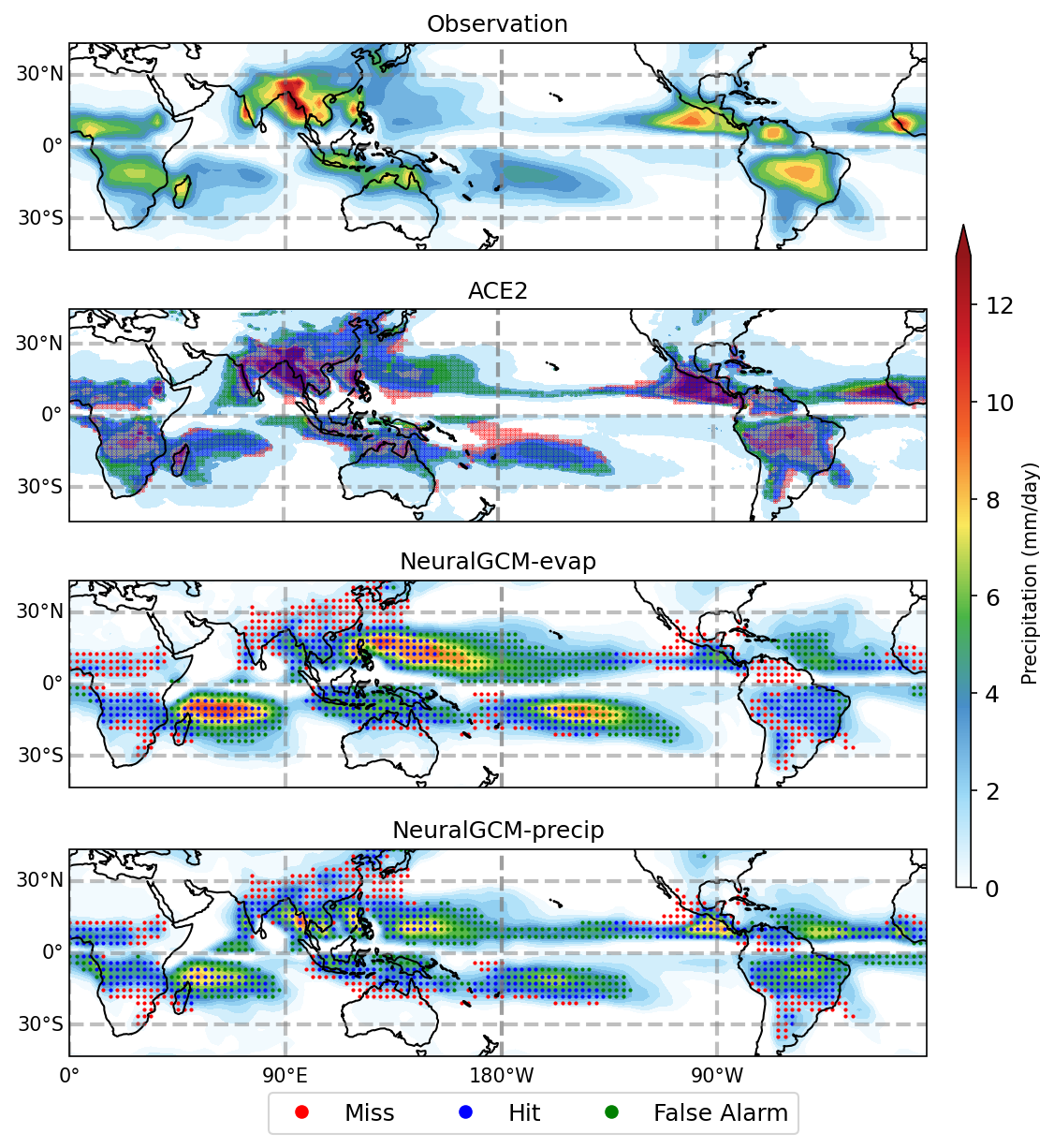}
    \caption{Annual precipitation range (shading, mm~day$^{-1}$) in regional monsoon domains for GPCP 3.2 observations (top; 1998--2017) and the three analyzed DL-ESMs. Blue stars indicate grid cells where the model and observations both meet the monsoon criterion, red dots indicate observed monsoon grid cells missed by the model, and green triangles indicate false alarms where the model meets the criterion but observations do not. The default threat-score threshold is 2.5~mm~day$^{-1}$.}
    \label{fig:monsoon_Wang}
\end{figure}

In addition to the temporal characterization, the spatial distribution of the annual precipitation range is evaluated using PMP metrics inspired by \citet{Wang2011}.
Figure~\ref{fig:monsoon_Wang} compares the observed and simulated annual spatial patterns of precipitation using a threshold-based classification of hits, misses, and false alarms.
All three DL-ESMs reproduce broad monsoonal structures, but ACE2 most closely captures the observed spatial pattern.
ACE2 also has a denser set of grid points because of its higher horizontal resolution.
The two NeuralGCM precipitation-capable variants capture several large-scale monsoon regions, including Northern Australia, parts of Africa, and South America, but they miss important observed monsoon behavior over India and parts of South and East Asia.
They also produce more widespread false alarms across the tropical oceans, indicating that the spatial placement of monsoon-like precipitation variability remains a limitation.

%---------------------- FIGURE 10 --------------------------

\subsection{Precipitation Variability Metrics} 
Accurate simulation of precipitation variability remains one of the most demanding tests for both physics-based and deep-learning models, because precipitation variability reflects interactions among moist convection, circulation, land--ocean contrast, and tropical waves across multiple timescales.
We therefore use PMP precipitation-variability metrics to evaluate not only whether the models reproduce the mean rainfall distribution, but also whether they capture the amplitude and geographic structure of rainfall fluctuations.

In PMP, precipitation variability is characterized using the simulated-to-observed ratio of spectral power, which reduces sensitivity to choices made in spectral processing, such as window length, overlap, and windowing function \citep{Ahn2022}.
A perfect model would have a ratio of 1. Values greater than 1 indicate excessive variability at a given timescale with respect to observations, whereas values below 1 indicate insufficient variability.
The spectral diagnostics separate total variability from anomaly variability, where the anomaly is obtained after removing the long-term mean and seasonal cycle.

%---------------------- FIGURE 11 --------------------------
\begin{figure}[h]
    \centering
    \includegraphics[width=0.9\textwidth]{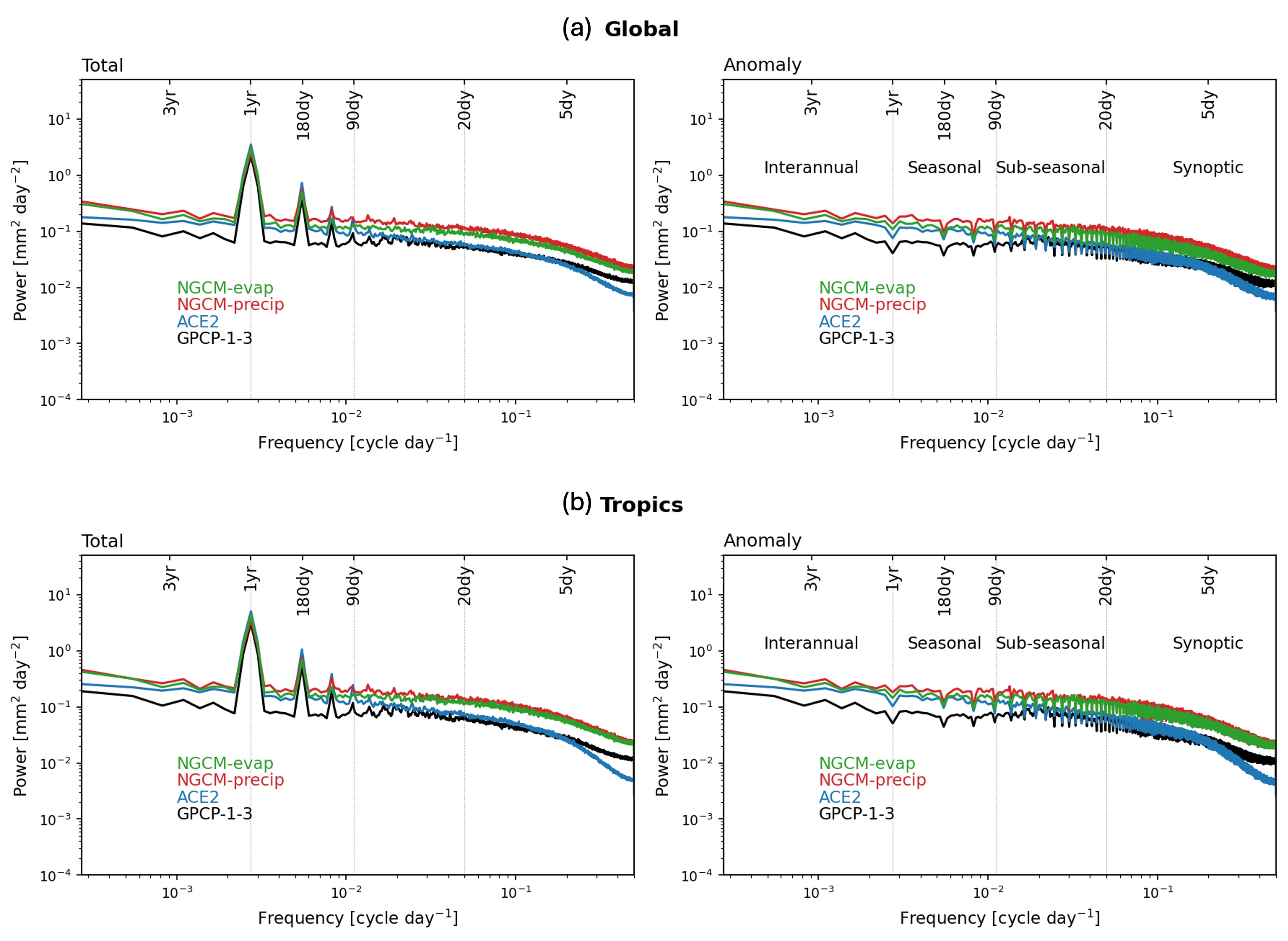}
    \caption{Power spectra of precipitation variability across timescales for the global domain (top row) and tropical domain (bottom row). Left panels show total precipitation variability, including the mean seasonal cycle, and right panels show anomaly variability after removing the long-term mean and seasonal cycle. The GPCP 1.3 reference is shown in black, and the DL-ESMs are shown by colored curves.}
    \label{fig:precip_var}
\end{figure}
%---------------------- --------------------------
% note : NeuralGCM: Often focuses on long-term climate stability and, in some versions, does not fully capture high-frequency Kelvin waves, while ACE2 (specifically ACE2-ERA5) can better capture certain high-frequency features as discussed in \citep{Baxter2026}
Figure~\ref{fig:precip_var} compares precipitation power spectra from ACE2, NeuralGCM-evap, and NeuralGCM-precip with the GPCP 1.3 reference for both global and tropical domains.
A realistic model should reproduce both the overall spectral slope and the peaks associated with annual, seasonal, sub-seasonal, and synoptic variability.
ACE2 follows the observed spectrum most closely across much of the frequency range, particularly for periods shorter than about one year, although it tends to overestimate power at interannual, seasonal, and sub-seasonal timescales in both the global and tropical domains.
NeuralGCM-evap and NeuralGCM-precip have broadly similar total spectra and generally remain farther from the GPCP reference than ACE2.
In the anomaly spectra, NeuralGCM-precip produces particularly large power over a wide range of timescales, indicating that direct precipitation prediction alone does not guarantee realistic variability.
However, the NeuralGCM-precip precipitation/convection formulation appears to improve behavior at synoptic timescales relative to longer-period variability.
In the sub-seasonal range ($\sim$20--90 days), ACE2 captures the MJO-relevant power more realistically, consistent with recent ACE2 analyses \citep{watt2025ace2}.

\begin{figure}[H]
    \centering
    \includegraphics[width=0.9\textwidth]{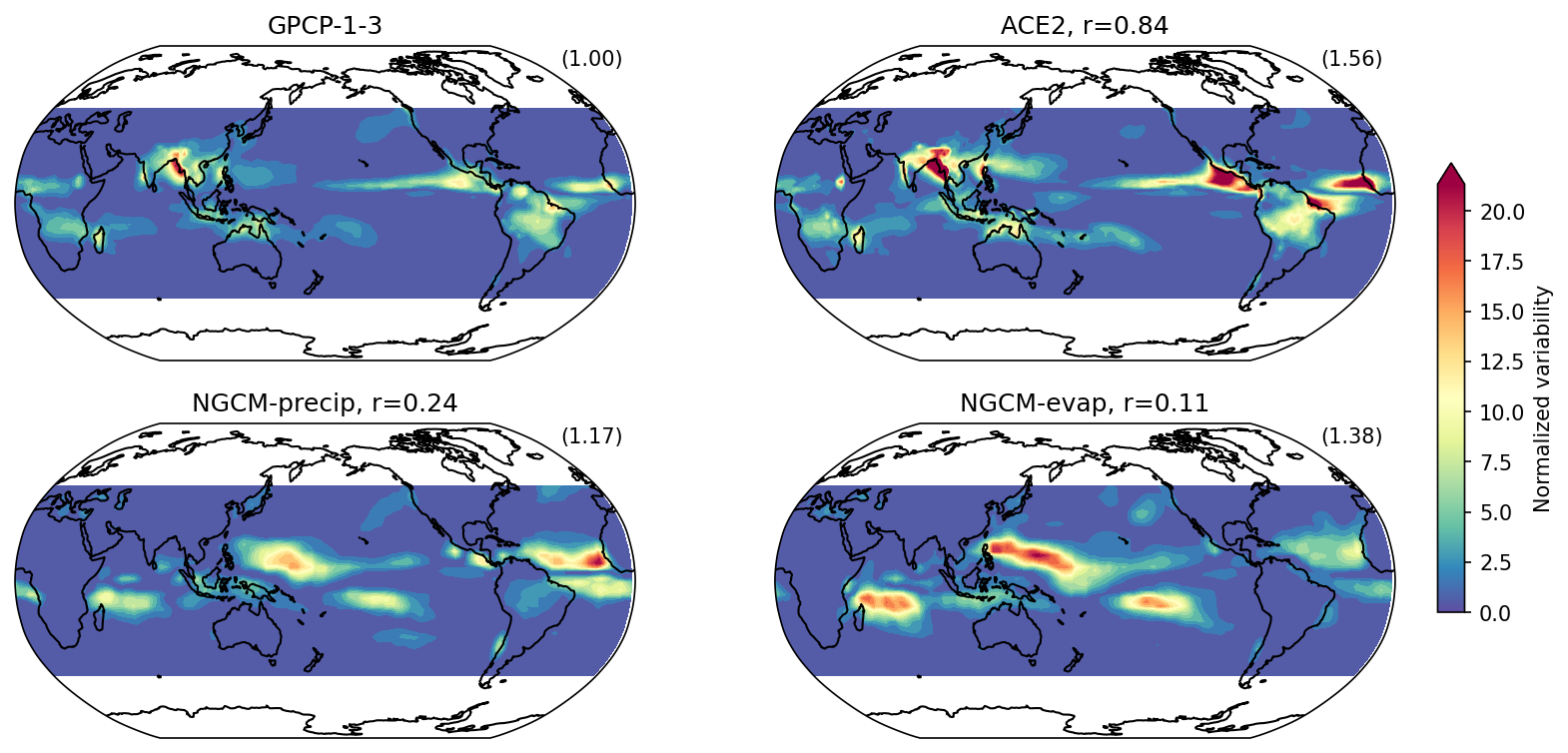}
    \caption{Normalized annual precipitation variability relative to the GPCP reference. Panel titles show the spatial pattern correlation ($r$) with observations, and values in parentheses show the ratio of model to observed spatial standard deviation. Values greater than 1 indicate excessive spatial variability amplitude relative to observations.}
    \label{fig:precip_annual}
\end{figure}

Figure~\ref{fig:precip_annual} maps normalized annual precipitation variability and provides a complementary spatial view of the spectral results.
The numbers in parentheses indicate the ratio of the model spatial standard deviation to the observed spatial standard deviation, while the panel titles report the spatial pattern correlation with GPCP 1.3.
ACE2 has the largest amplitude ratio (1.56), meaning that its annual variability is too strong overall, but it also has the highest spatial correlation with observations ($r=0.84$).
This combination indicates that ACE2 captures the main regions of annual precipitation variability while overestimating their amplitude.
For example, ACE2 reproduces enhanced variability in the Indian monsoon and tropical convergence-zone regions that are largely absent or displaced in the two NeuralGCM variants.
In contrast, NeuralGCM-precip and NeuralGCM-evap have lower amplitude ratios (1.17 and 1.38, respectively) but considerably lower spatial correlations ($r=0.24$ and $r=0.11$), showing that an apparently reasonable global amplitude can mask poor geographic placement.
Both NeuralGCM variants emphasize variability over the western and central tropical Pacific and miss several observed high-variability regions.
The semi-annual and interannual maps in Figs.~\ref{SIfig:precip_semiannual} and~\ref{SIfig:precip_inter} show similar behavior, reinforcing that precipitation variability must be evaluated with both amplitude and spatial-pattern metrics.

%---------------------- FIGURE 12 --------------------------
% NEW
\subsection{Taylor Diagrams}
\begin{figure}[H]
    \centering
    \includegraphics[width=0.99\textwidth]{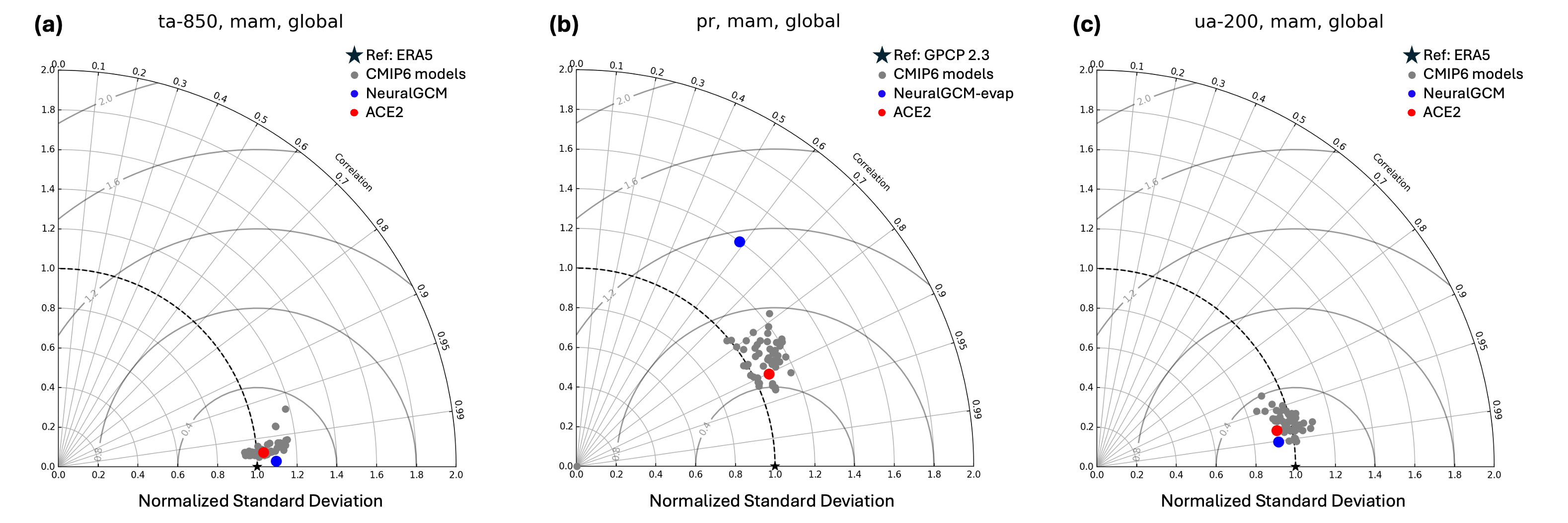}
    \caption{Taylor diagrams comparing DL-ESMs with CMIP6 models for March--May global fields: (a) air temperature at 850~hPa, (b) precipitation, and (c) zonal wind at 200~hPa. Diagrams show normalized standard deviation, spatial pattern correlation, and centered RMSE relative to the corresponding reference dataset.}
    \label{fig:taylor_3}
\end{figure}

Taylor diagrams \citep{Taylor2001} summarize normalized standard deviation, spatial pattern correlation, and RMSE in a single performance space.
Within the PMP workflow, they provide a compact benchmark for assessing whether the DL-ESMs occupy the same performance range as CMIP6 models and how closely they approach the observational or reanalysis reference for individual variables and seasons \citep{lee2024PMP}.
Figure~\ref{fig:taylor_3} illustrates this PMP tool and compares March--May (MAM) global fields for 850~hPa air temperature, precipitation, and 200~hPa zonal wind.

For 850-hPa air temperature (Fig.~\ref{fig:taylor_3}a), ACE2 lies very close to the ERA5 reference point, with a normalized standard deviation near unity and an almost perfect spatial pattern correlation.
NeuralGCM also preserves the broad MAM temperature pattern but is slightly farther from the reference.
Thus, the Taylor diagram indicates that the DL-ESMs do not behave identically for lower-tropospheric temperature: NeuralGCM is comparable to the best-performing CMIP6 models, whereas ACE2 lies closer to the lower-skill edge of the CMIP6 distribution for this field. The precipitation panel (Fig.~\ref{fig:taylor_3}b) is the most discriminating of the three diagnostics.
Relative to GPCP 2.3, ACE2 lies close to the reference point, with a normalized standard deviation close to one and a high spatial pattern correlation.
This placement is consistent with the precipitation maps and variability diagnostics above, which show that ACE2 captures much of the large-scale MAM precipitation structure despite regional biases.
In contrast, NeuralGCM-evap has a substantially larger normalized standard deviation, roughly 1.2, and a lower spatial pattern correlation, roughly 0.6.
Its larger angular separation from the reference point gives it a larger centered RMSE, indicating that the main error is not only excessive precipitation variability but also the geographic organization of precipitation.
The CMIP6 models form a broad cloud between these two DL-ESM behaviors, highlighting the well-known difficulty of reproducing global precipitation patterns and showing that ACE2 performs unusually well for this particular MAM precipitation metric, while NeuralGCM-evap remains outside the better-performing part of the CMIP6 distribution.

For 200-hPa zonal wind (Fig.~\ref{fig:taylor_3}c), both DL-ESMs fall within the CMIP6 spread and close to ERA5.
NeuralGCM is slightly closer to the reference, while ACE2 has similarly high pattern correlation but somewhat weaker normalized variability.
Taken together, the Taylor diagrams show that the DL-ESMs can reproduce large-scale dynamical and thermodynamic fields with CMIP6-like skill, but that precipitation remains a more stringent test: ACE2 is close to the observational reference for MAM global precipitation, whereas NeuralGCM-evap exhibits larger pattern and amplitude errors.

%---------------------- FIGURE 13 --------------------------
%------------------------------------------------

\section{Discussion and Conclusions}
\label{Sect:disc}

This study evaluates ACE2, NeuralGCM, NeuralGCM-evap, and NeuralGCM-precip through the lens of PMP diagnostics traditionally used for CMIP-class Earth System models.
The selected metrics cover complementary aspects of Earth system performance, including mean climatology, modes of variability, monsoon behavior, and precipitation variability.
By applying the same framework to DL-ESMs and physics-based CMIP models, we show that PMP provides a practical and reproducible way to identify where DL-ESMs are already competitive and where they require further development.

The results indicate that several large-scale atmospheric fields are represented with encouraging skill.
Temperature and wind fields in the DL-ESMs are often comparable to, and in some cases better than, the CMIP6 benchmark distribution for the historical-period diagnostics considered here.
This performance is strongest for variables that are directly available to the models during training or appear explicitly in their objective functions.

The AMIP-style experimental design nevertheless remains a meaningful test because the simulations use prescribed boundary-condition datasets and time-varying forcings, and the evaluation periods are not always identical to the training configuration. Thus, even though SST and sea-ice conditions are imposed, the models must still respond realistically to evolving boundary conditions and external forcing trajectories. This is also relevant for the next generation of AI-based Earth system models, which are likely to incorporate a broader range of time-varying forcings and to be increasingly coupled to additional Earth system components, including ocean, land and sea ice modules. The PMP framework is therefore well suited not only for evaluating the AMIP-style DL-ESMs considered here, but also for testing future coupled and forced AI-based ESMs under more comprehensive climate-modeling configurations.

Precipitation is the clearest limitation and the most informative stress test in this analysis.
The mean precipitation maps, spectral diagnostics, annual-variability maps, monsoon diagnostics, and Taylor diagrams all show that precipitation errors differ substantially across DL-ESM formulations.
ACE2 captures the broad spatial organization of precipitation and precipitation variability better than the NeuralGCM variants considered here, including realistic high-variability regions associated with tropical convergence zones and monsoon rainfall.
At the same time, ACE2 tends to over-intensify tropical precipitation and overestimate variability amplitude, indicating that good pattern skill does not imply a fully correct hydrological cycle.
NeuralGCM-evap shows widespread tropical wet biases and lower spatial correlations, consistent with the difficulty of deriving precipitation indirectly from precipitation minus evaporation.
NeuralGCM-precip improves some high-frequency variability but also exhibits anomalous climatology and stability limitations.
Together, these results suggest that DL-ESM precipitation evaluation must combine mean-state, spectral, spatial-pattern, and regional-process diagnostics rather than relying on a single global error metric.

The stability analysis further distinguishes the models.
ACE2 remains stable over the long roll-outs considered here, whereas NeuralGCM-evap and NeuralGCM-precip show greater sensitivity to initial conditions, random seeds, and run length.
The original NeuralGCM at $2.8^\circ$ resolution remains stable across the ten deterministic ensemble members when the global mean logarithm of surface pressure is conserved, following the correction provided by the NeuralGCM developers.
However, the precipitation-enabled NeuralGCM versions do not have the same ability to complete a prescribed run length.
This behavior is not unexpected for a hybrid model, which can inherit stability challenges similar to those encountered in physics-based ESMs \citep{Kochkov2024}, but it is especially important for Earth system science applications that require multi-year or multi-decadal integrations.

These findings point to several development priorities.
For precipitation, improvements are needed not only in mean rainfall amount but also in the representation of tropical convergence-zone structure, monsoon timing, and variability across synoptic-to-interannual timescales.
Because tropical precipitation biases are tied to small-scale moist processes and their feedbacks, higher resolution alone is unlikely to remove all errors.
For DL-ESMs, progress may require changes in training targets, loss functions, conservation constraints, stochastic parameterizations, or architectures that better connect precipitation to circulation, moisture, and surface-flux processes.
The PMP diagnostics used here provide a way to track whether such changes improve the intended process without degrading other aspects of the simulated Earth system.

This study supports the use of PMP-style diagnostics for extending DL-ESM assessment beyond weather forecast lead times toward subseasonal-to-seasonal and Earth system science applications.
Metrics targeting modes of variability such as the Madden--Julian Oscillation, monsoon onset and decay, and precipitation variability are particularly relevant because they connect short-timescale weather processes to longer-timescale Earth system behavior.
Future work should expand the analysis to additional DL-ESMs, including foundation models such as Aurora \citep{Aurora_2025} and coupled systems such as DLESyM \citep{DLESyM_2025}, and should further evaluate extremes, including storyline applications such as the NeuralGCM heatwave analysis of \citet{Duan2025}.

Overall, our results confirm the potential of DL-ESMs as computationally efficient complements to traditional ESMs while also clarifying that their fitness for purpose is variable-, process-, and timescale-dependent.
A standardized diagnostic framework such as PMP can therefore help the community move from general claims about DL-ESM skill toward transparent, process-specific benchmarks that guide model development and support emerging efforts such as AIMIP.
% \appendix  

\authorcontribution{S.D., G.P., C.B., J.L. and P.U. designed the experiments, formulated the article structure and contributed to the interpretation of the results. S.D. conducted the experiments. S.D., S.G. and J.L. produced the analysis. All authors analyzed the results and reviewed the manuscript.} %% this section is mandatory

\competinginterests{One of the coauthors is a member of the editorial board of Geoscientific Model Development. The peer-review process was guided by an independent editor, and the authors also have no other competing interests to declare.} %% this section is mandatory even if you declare that no competing interests are present

% \disclaimer{TEXT} %% optional section
\section*{Code and data availability}

The PMP source code is available as an open-source Python package at 
\url{https://github.com/PCMDI/pcmdi_metrics}, with all released versions archived on Zenodo 
(\url{https://doi.org/10.5281/zenodo.592790}; \citealp{lee2024PMP}). 
Documentation is provided at \url{http://pcmdi.github.io/pcmdi_metrics}.
The PMP results database, including calculated metrics, is available at 
\url{https://github.com/PCMDI/pcmdi_metrics_results_archive}, with archived versions on Zenodo 
(\url{https://doi.org/10.5281/zenodo.10181201}; \citealp{lee2024PMP}).
The PMP can be installed via the Anaconda distribution using the conda-forge channel 
(\url{https://anaconda.org/conda-forge/pcmdi_metrics}), with installation instructions available at 
\url{http://pcmdi.github.io/pcmdi_metrics/install.html}.
Interactive visualizations of PMP results are available at 
\url{https://pcmdi.llnl.gov/metrics}.
The CMIP6 model outputs and obs4MIPs datasets used in this study are accessible via the 
Earth System Grid Federation (ESGF; \url{https://esgf-node.llnl.gov/}). The model checkpoints for ACE2 and NeuralGCM are available at \url{https://huggingface.co/allenai/ACE2-ERA5} and \url{https://neuralgcm.readthedocs.io/en/latest/}, respectively. 

\begin{acknowledgements}
This work was performed under the auspices of the U.S. Department of Energy by Lawrence Livermore National Laboratory under Contract DE-AC52-07NA27344. 
Review record LLNL-JRNL-2017669.
G.P., S.D., C.B., J.L. and P.U. are funded by PCMDI Science Focus Area. 
The authors would like to acknowledge the help from 
Dr. Min-Seop Ahn for precipitation variability and Dr. Peter Gleckler for CMORization.
The authors used AI-based writing tools only for copyediting and language polishing. All scientific content, analyses, interpretations, and final text were reviewed and approved by the authors.
\end{acknowledgements}

%% REFERENCES
% \bibliographystyle{plainnat}
\bibliographystyle{copernicus}
\bibliography{references_PMP_DLESM}

\section{Methods}
% \subsection{Boundary Conditions }
% This data set combines two specific SST and SI data sets: Hadley Centre Sea Ice and SST (HadISST) version 1.1 through 1981, and Optimum Interpolation Sea Surface Temperature (OISST) from the National Oceanic and Atmospheric Administration (NOAA) version 2.0 after then (Hurrell et al., 2008).
\subsection{Standardizing Data and Model Output}
\begin{itemize}
    \item \textbf{Data Formatting and Preprocessing:}  
    NeuralGCM and ACE2 outputs were post-processed to meet common data formats. This involves regridding or interpolating simulated fields (e.g., temperature, wind components) onto a common pressure level (e.g., 500 hPa, 200 hPa) for ACE2. In addition to the default outputs, sea level pressure is approximated from surface pressure and topography. The Climate Model Output Rewriter (CMOR) are used to process the model simulations following the CF conventions. 

    \item \textbf{Reference Datasets:}  
    For model evaluation, we selected state-of-the-art observational products (such as ERA5 reanalysis for atmospheric fields, GPCP for precipitation, etc.) that match the spatial and temporal scopes of NeuralGCM and ACE2 simulations. The reference datasets are also processed in a CF-compliant format. Rigorous intercomparison between the AI model and observations requires that variable definitions and units are consistent.
\end{itemize}

\section*{Appendix A: Acronyms}

\begin{longtable}{ll}
\tophline
Acronym & Description \\
\middlehline
AMIP & Atmospheric Model Intercomparison Project \\
CMIP & Coupled Model Intercomparison Project \\
% CMOR & Climate Model Output Rewriter \\
% CVDP & Climate Variability Diagnostics Package \\
% DOE & U.S. Department of Energy \\
ENSO & El Niño–Southern Oscillation \\
EOF & Empirical orthogonal function \\
EOR & East power normalized by observation \\
% ESGF & Earth System Grid Federation \\
ESM & Earth system model \\
% ESMAC Diags & Earth System Model Aerosol–Cloud Diagnostics \\
ETMoV & Extratropical modes of variability \\
EWR & East–west power ratio \\
GFDL & Geophysical Fluid Dynamics Laboratory \\
MAE & Mean absolute error \\
MJO & Madden–Julian Oscillation \\
NAM & Northern Annular Mode \\
NAO & North Atlantic Oscillation \\
NASA & National Aeronautics and Space Administration \\
NPO & North Pacific Oscillation \\
PCMDI & Earth System Model Evaluation Project \\
PDO & Pacific Decadal Oscillation \\
PMP & PCMDI Metrics Package \\
PNA & Pacific North America pattern \\
RMSE & Root Mean Square Error \\
SAM & Southern Annular Mode \\
SH & Southern Hemisphere \\
SST & Sea Surface Temperature \\
TOA & Top of Atmosphere \\
\bottomhline
\end{longtable}

% Table~\ref{tab:mov} shows the main Modes of Variability addressed in our study.
% \begin{table}[h]
% \caption{Overview of the investigated modes of variability in PMP framework (as in \citep{lee2024PMP}).}
% % \begin{tabular}{column = lc}
% \small
% \begin{tabular}{ll}
% \tophline
% Mode of Variability & Description \\
% \middlehline
% MJO & Madden–Julian Oscillation \\
% NAM & Northern Annular Mode \\
% NAO & North Atlantic Oscillation \\
% NPO & North Pacific Oscillation \\
% PDO & Pacific Decadal Oscillation \\
% PNA & Pacific North America pattern \\
% SAM & Southern Annular Mode \\
% \bottomhline
% \end{tabular}
% % \belowtable{} % Table Footnotes
% \label{tab:mov}
% \end{table}

%%%%%%%%%%%%%%%%%%%%%%%% SI FIGURES  %%%%%%%%%%%%%%%%%%%%%%%%
\clearpage
\newpage
\setcounter{figure}{0}
\input{1_Supplementary_material_revised}

\end{document}

%% file: 1_Supplementary_material_revised.tex
%%%%%%%%%%%%%%%%%%%%%%%% SI FIGURES  %%%%%%%%%%%%%%%%%%%%%%%%
\clearpage
\newpage
\setcounter{figure}{0}
\clearpage
% \appendix
\section{Supplementary Figures}
\setcounter{figure}{0}
\renewcommand{\thefigure}{S\arabic{figure}}

This supplement provides additional diagnostic figures supporting the PMP-based evaluation of the deep-learning Earth system models (DL-ESMs) discussed in the main manuscript. The figures extend the main text by showing individual climatological fields, complementary portrait-plot diagnostics, Madden--Julian Oscillation (MJO) propagation metrics, regional monsoon timing diagnostics, and additional precipitation-variability timescales. Unless otherwise noted, the diagnostics use the standardized model output and reference datasets described in the Methods section and in Tables~\ref{tab:models} and \ref{tab:var}.

\subsection{Climatology}
\label{SI:clim}

The climatology figures summarize annual and seasonal mean fields over the common historical evaluation period. The variables include precipitation (pr), sea-level pressure (psl), air temperature (ta), zonal wind (ua), meridional wind (va), and geopotential height (zg). Pressure-level variables are reported at the levels used in the main PMP climatology analysis, and precipitation is compared against the corresponding GPCP reference product. These maps provide the field-level context for the normalized RMSE and MAE summaries discussed in the main text.

%---------------------- FIGURE S1 --------------------------
\begin{figure}[h]
    \centering
    \includegraphics[width=0.5\textwidth]{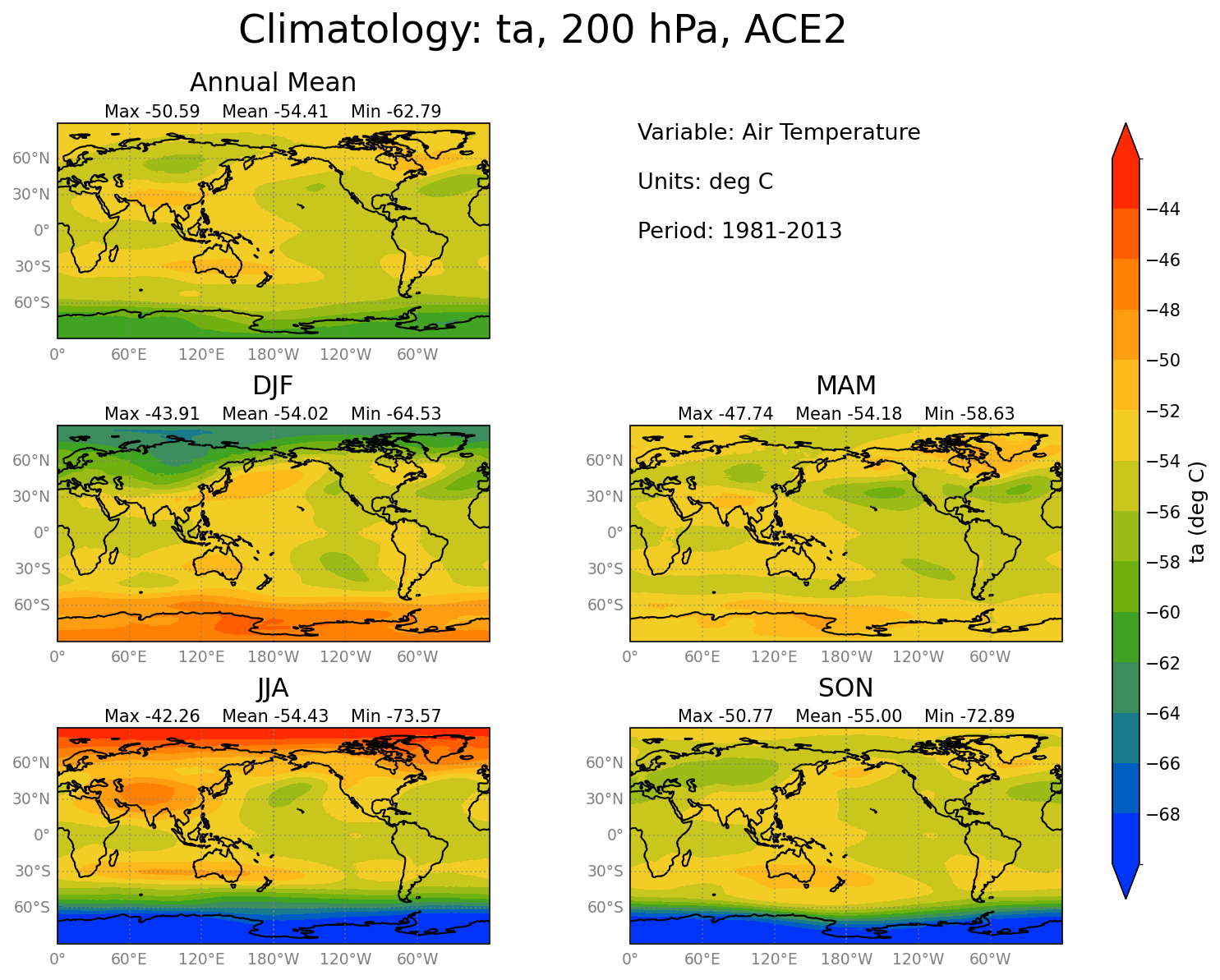}
    \caption{Mean air temperature at 200~hPa (ta-200) simulated by ACE2 for 1981--2013.}
    \label{SIfig:clim_ACE_ta200}
\end{figure}

%---------------------- FIGURE S2 --------------------------
\begin{figure}[h]
    \centering
    \includegraphics[width=0.4\textwidth]{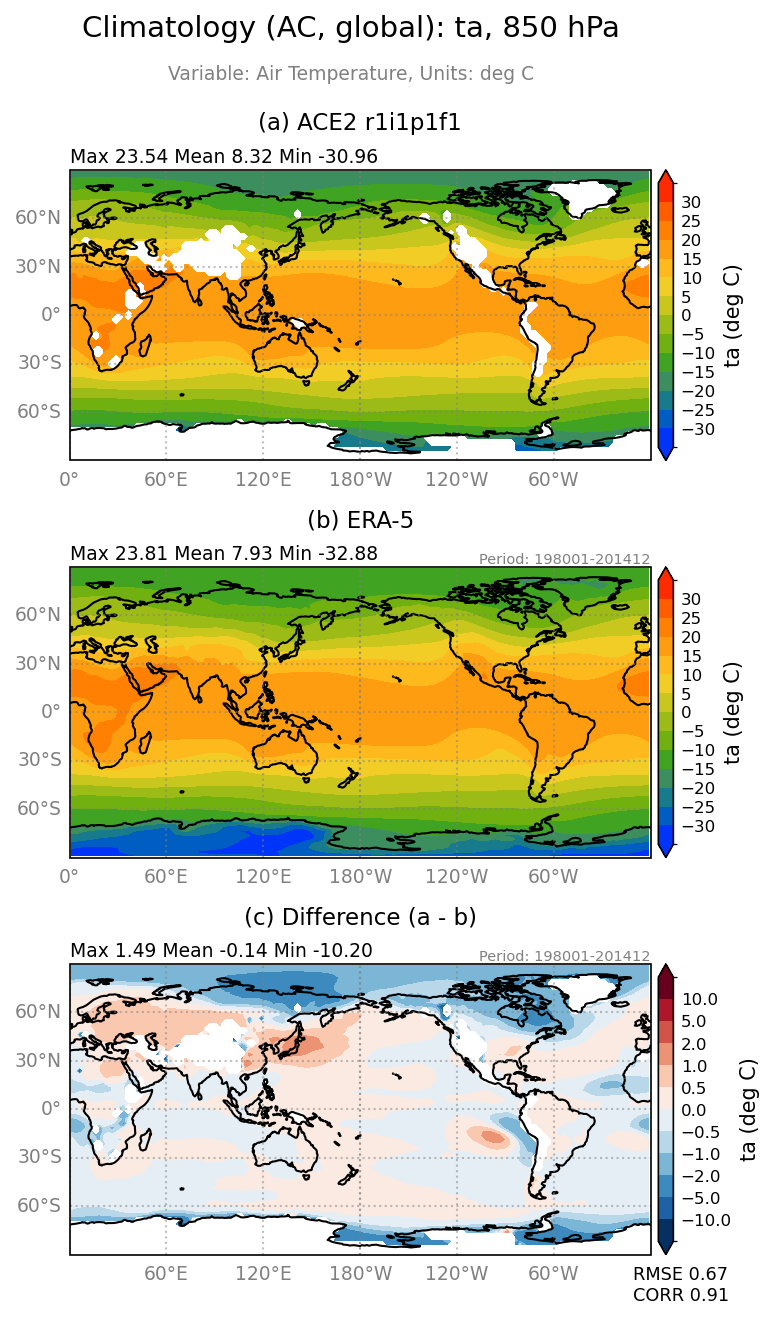}
    \caption{Annual-mean air temperature at 850~hPa (ta-850) from ACE2 compared with ERA5. The panels show (a) the ACE2 field, (b) the ERA5 reference field, and (c) the ACE2 minus ERA5 bias discussed in the main text.}
    \label{SIfig:ACEbias_ta850}
\end{figure}
%---------------------- FIGURE S3 --------------------------
\begin{figure}[h]
    \centering
    \includegraphics[width=0.5\textwidth]{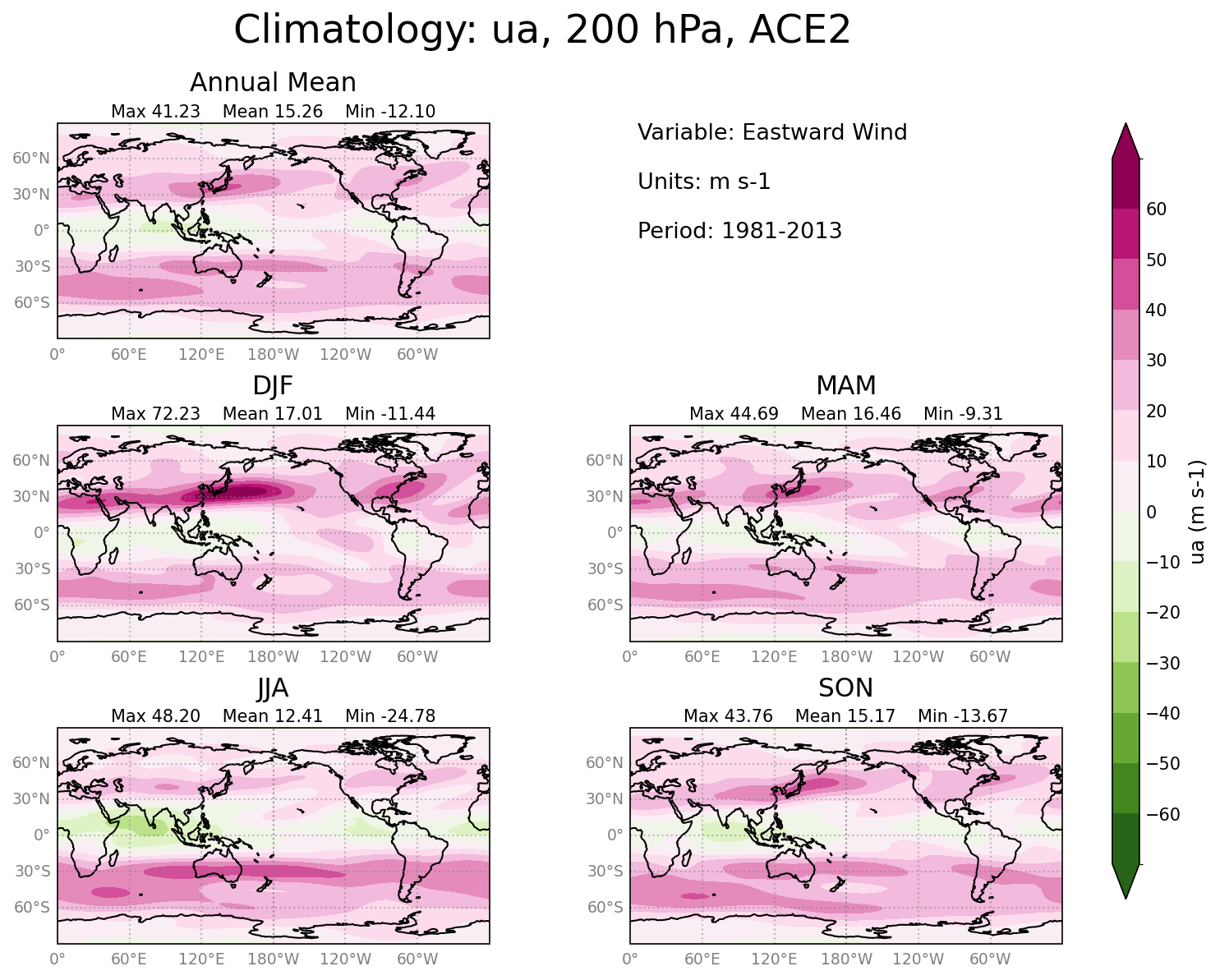}
    \caption{Mean zonal wind at 200~hPa (ua-200) simulated by ACE2 for 1981--2013. The annual and seasonal means are shown in the same panel layout as Fig.~\ref{SIfig:clim_ACE_ta200}.}
    \label{SIfig:clim_ACE_ua200}
\end{figure}
%---------------------- FIGURE S4 --------------------------
\begin{figure}[h]
    \centering
    \includegraphics[width=0.5\textwidth]{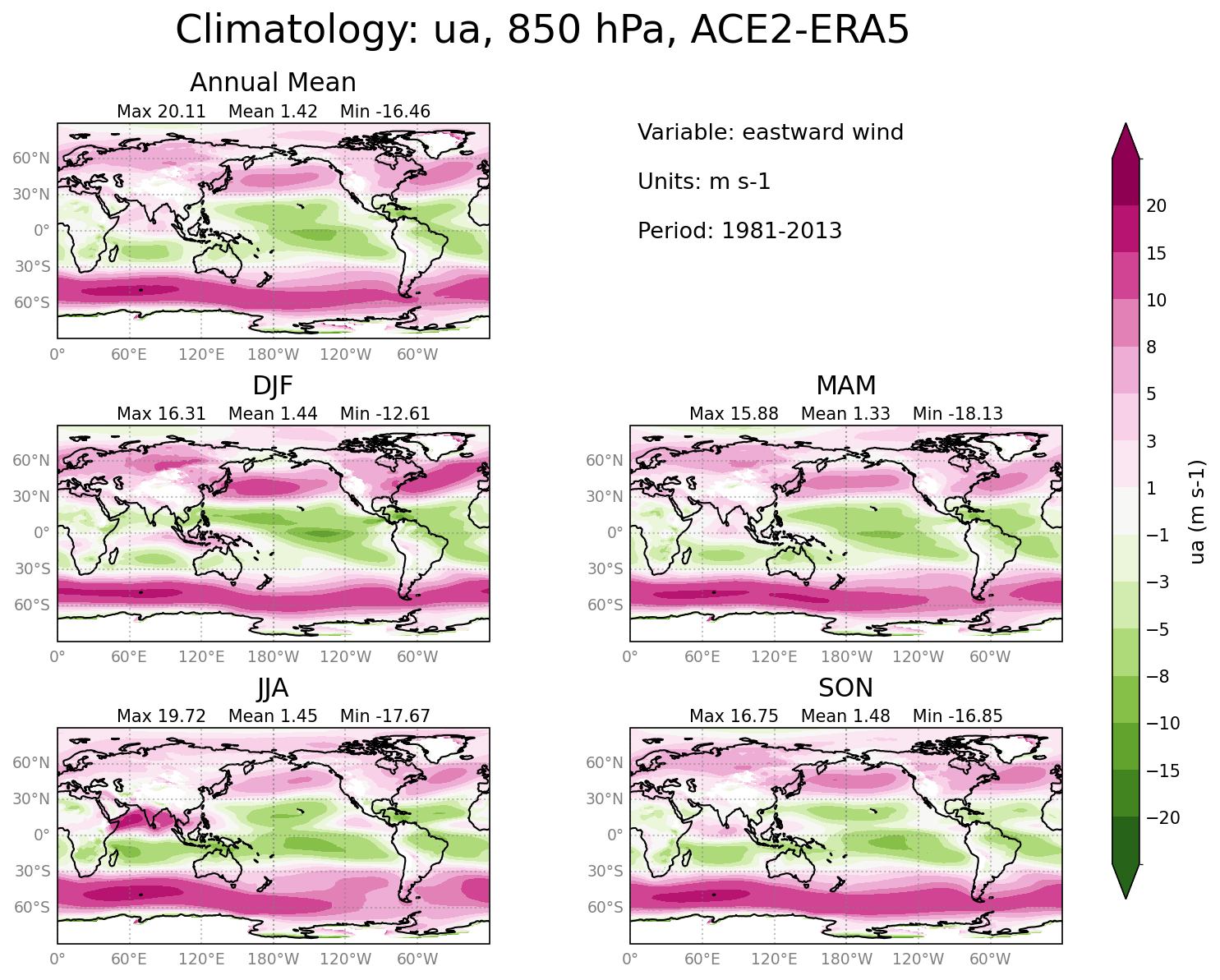}
    \caption{Mean zonal wind at 850~hPa (ua-850) simulated by ACE2 for 1981--2013. The annual and seasonal means are shown in the same panel layout as Fig.~\ref{SIfig:clim_ACE_ta200}.}
    \label{SIfig:clim_ACE_ua850}
\end{figure}

%---------------------- FIGURE S5 --------------------------
\begin{figure}[h]
    \centering
    \includegraphics[width=0.5\textwidth]{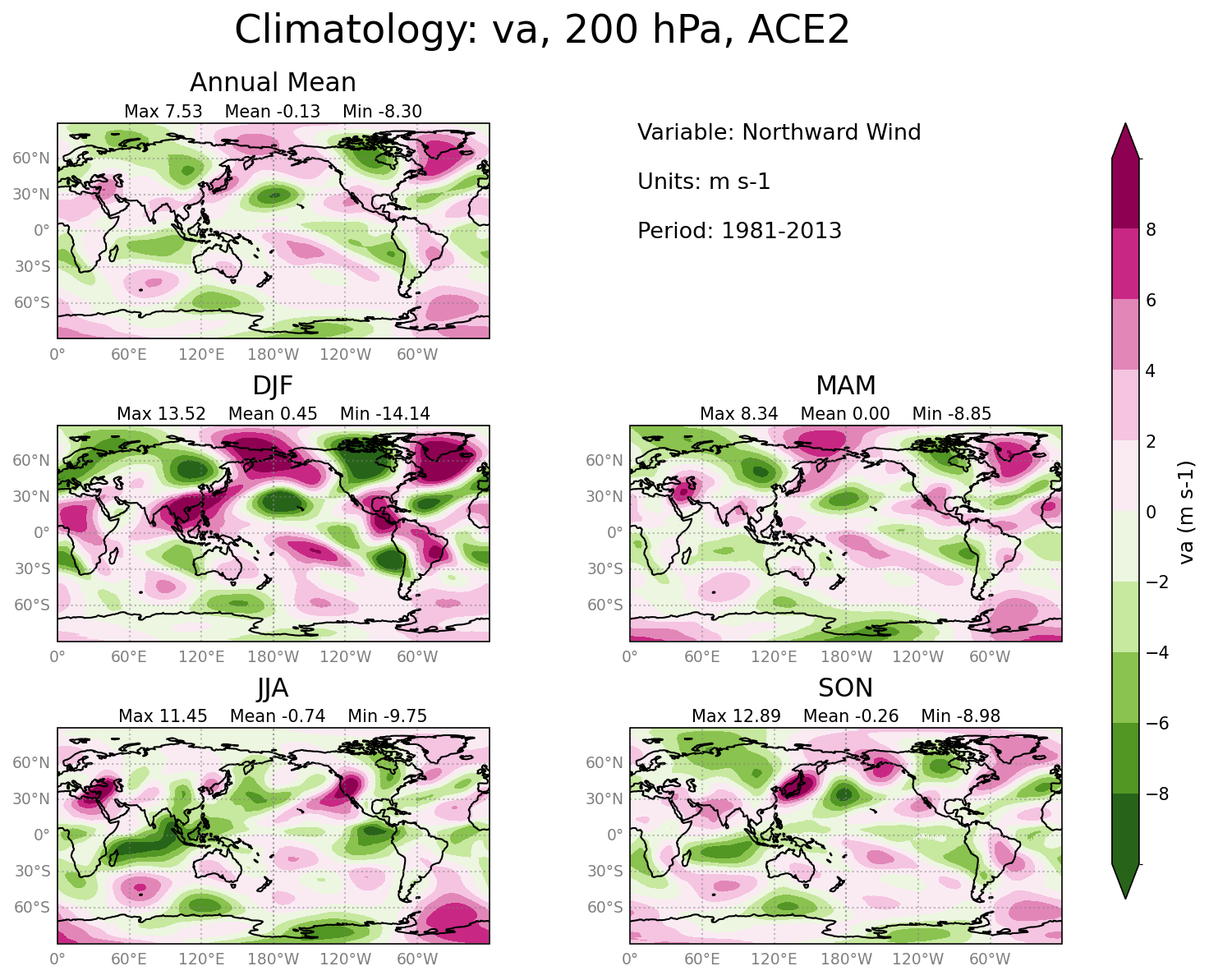}
    \caption{Mean meridional wind at 200~hPa (va-200) simulated by ACE2 for 1981--2013. The annual and seasonal means are shown in the same panel layout as Fig.~\ref{SIfig:clim_ACE_ta200}.}
    \label{SIfig:clim_ACE_va200}
\end{figure}

%---------------------- FIGURE S6 --------------------------
\begin{figure}[h]
    \centering
    \includegraphics[width=0.5\textwidth]{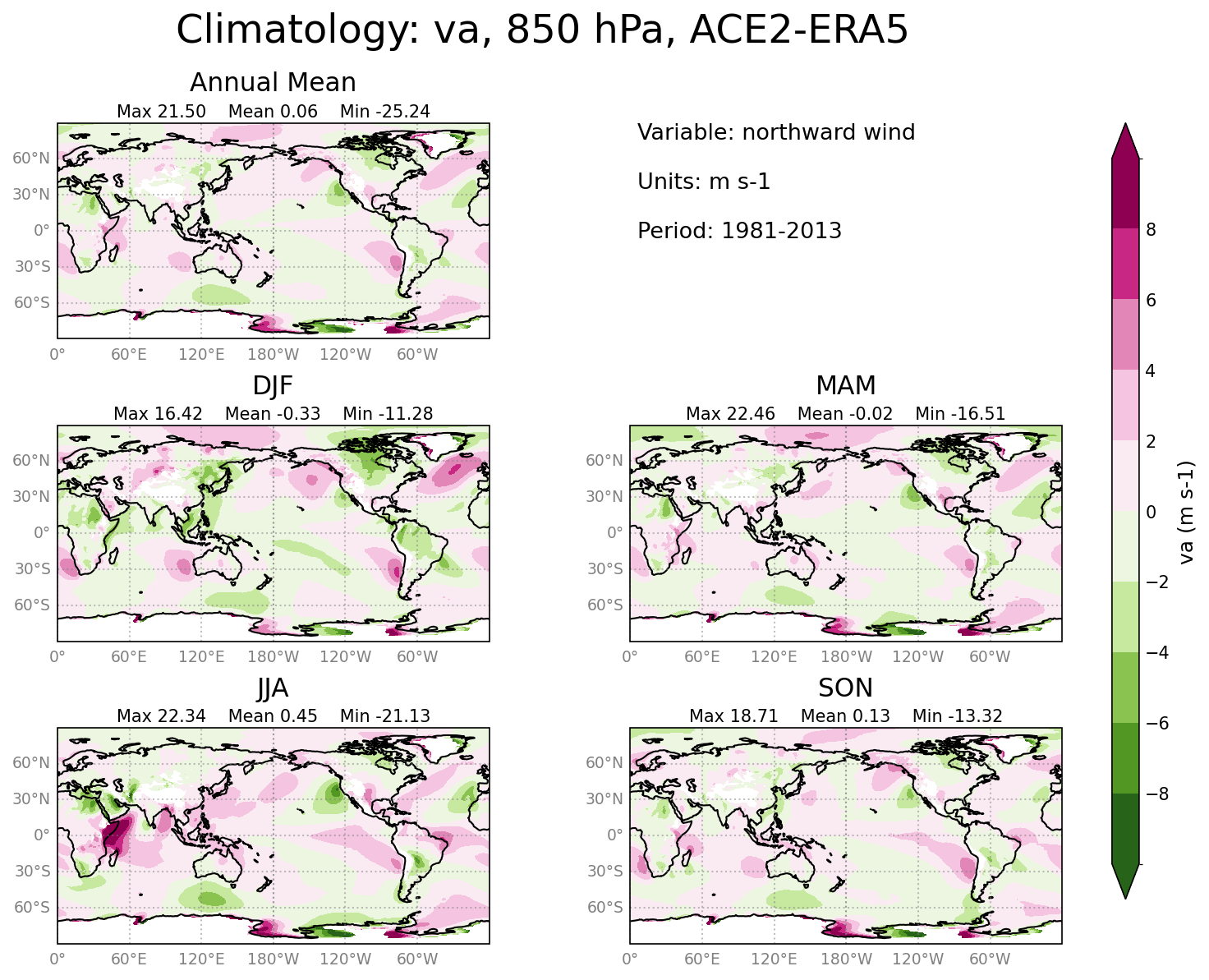}
    \caption{Mean meridional wind at 850~hPa (va-850) simulated by ACE2 for 1981--2013. The annual and seasonal means are shown in the same panel layout as Fig.~\ref{SIfig:clim_ACE_ta200}.}
    \label{SIfig:clim_ACE_va850}
\end{figure}
%---------------------- FIGURE S7 --------------------------

\begin{figure}[h]
    \centering
    \includegraphics[width=0.5\textwidth]{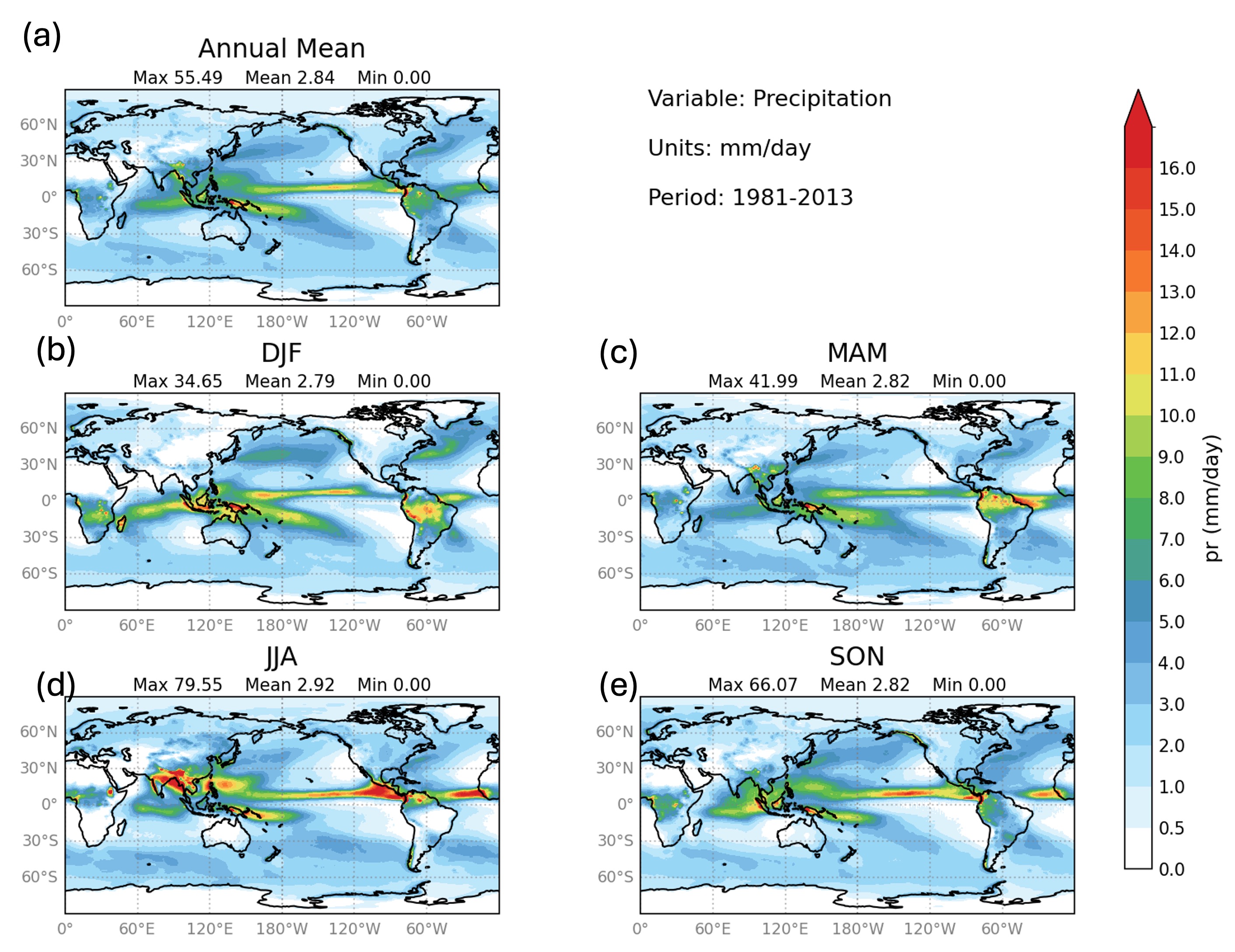}
    \caption{Mean precipitation simulated by ACE2 for 1981--2013. Panel (a) shows the annual mean, and panels (b)--(e) show seasonal means for DJF, MAM, JJA, and SON, respectively. Units are mm~day$^{-1}$.}
    \label{SIfig:clim_ACE_pr}
\end{figure}

%---------------------- FIGURE S8 --------------------------

\begin{figure}[h]
    \centering
    \includegraphics[width=0.5\textwidth]{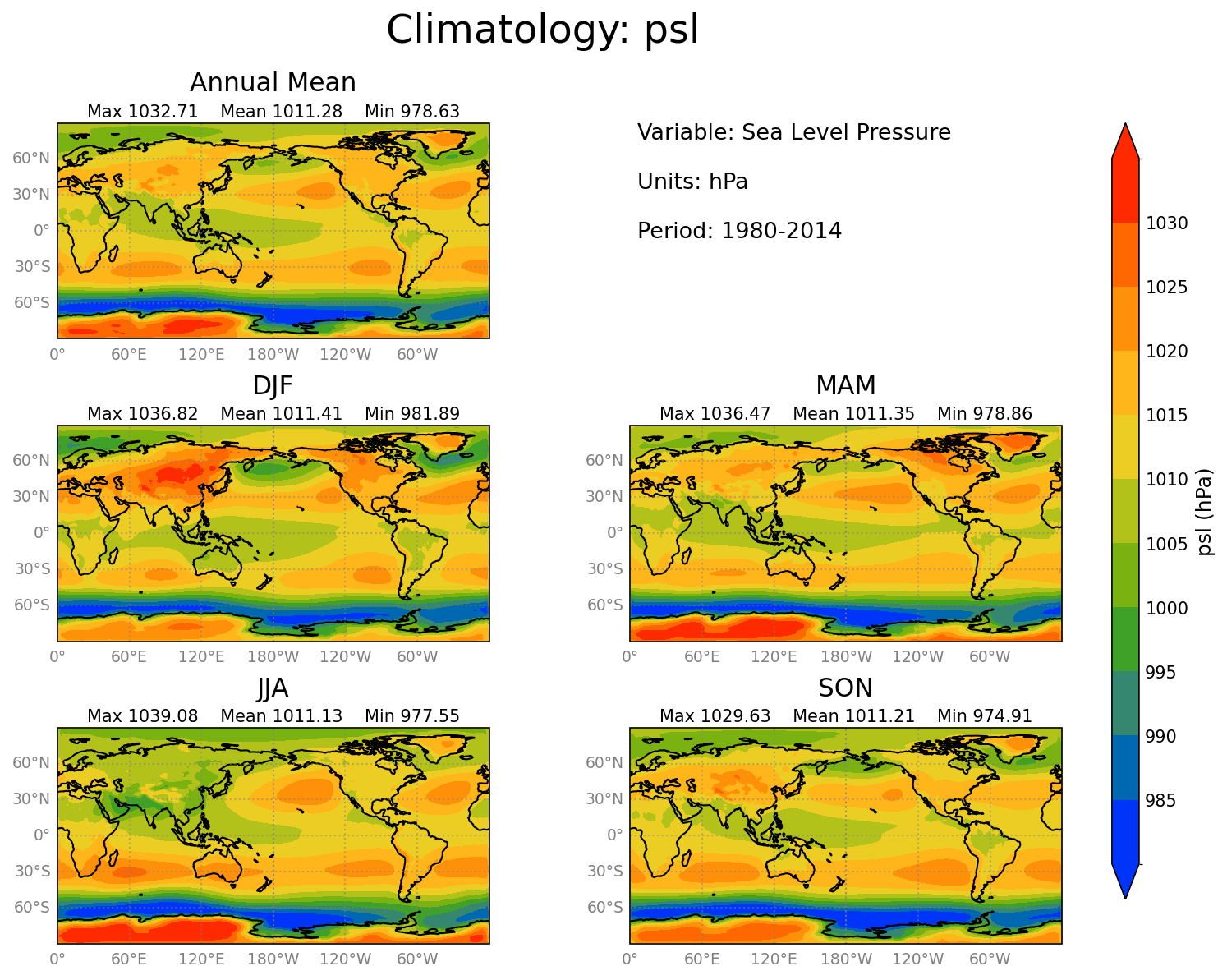}
    \caption{Mean sea-level pressure (psl) simulated by ACE2 for 1981--2013. Panel (a) shows the annual mean, and panels (b)--(e) show seasonal means for DJF, MAM, JJA, and SON, respectively. Units are hPa.}
    \label{SIfig:clim_ACE_psl}
\end{figure}

%---------------------- FIGURE S9 --------------------------
\begin{figure}[h]
    \centering
    \includegraphics[width=0.5\textwidth]{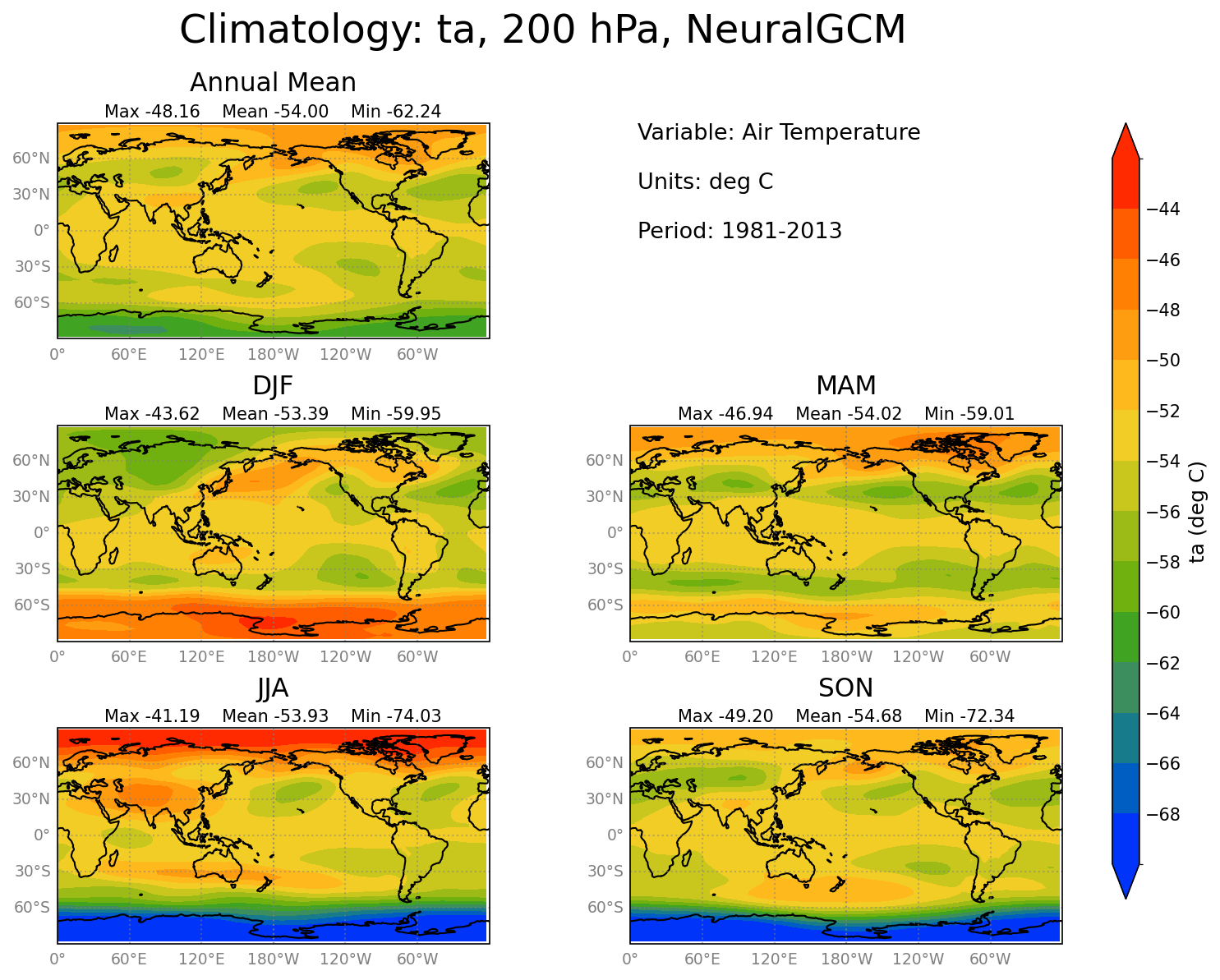}
    \caption{Mean air temperature at 200~hPa (ta-200) simulated by NeuralGCM for 1981--2013. The annual and seasonal means are shown in the same panel layout as Fig.~\ref{SIfig:clim_ACE_ta200}.}
    \label{SIfig:clim_NeuralGCM_ta200}
\end{figure}

%---------------------- FIGURE S10 --------------------------
\begin{figure}[h]
    \centering
    \includegraphics[width=0.5\textwidth]{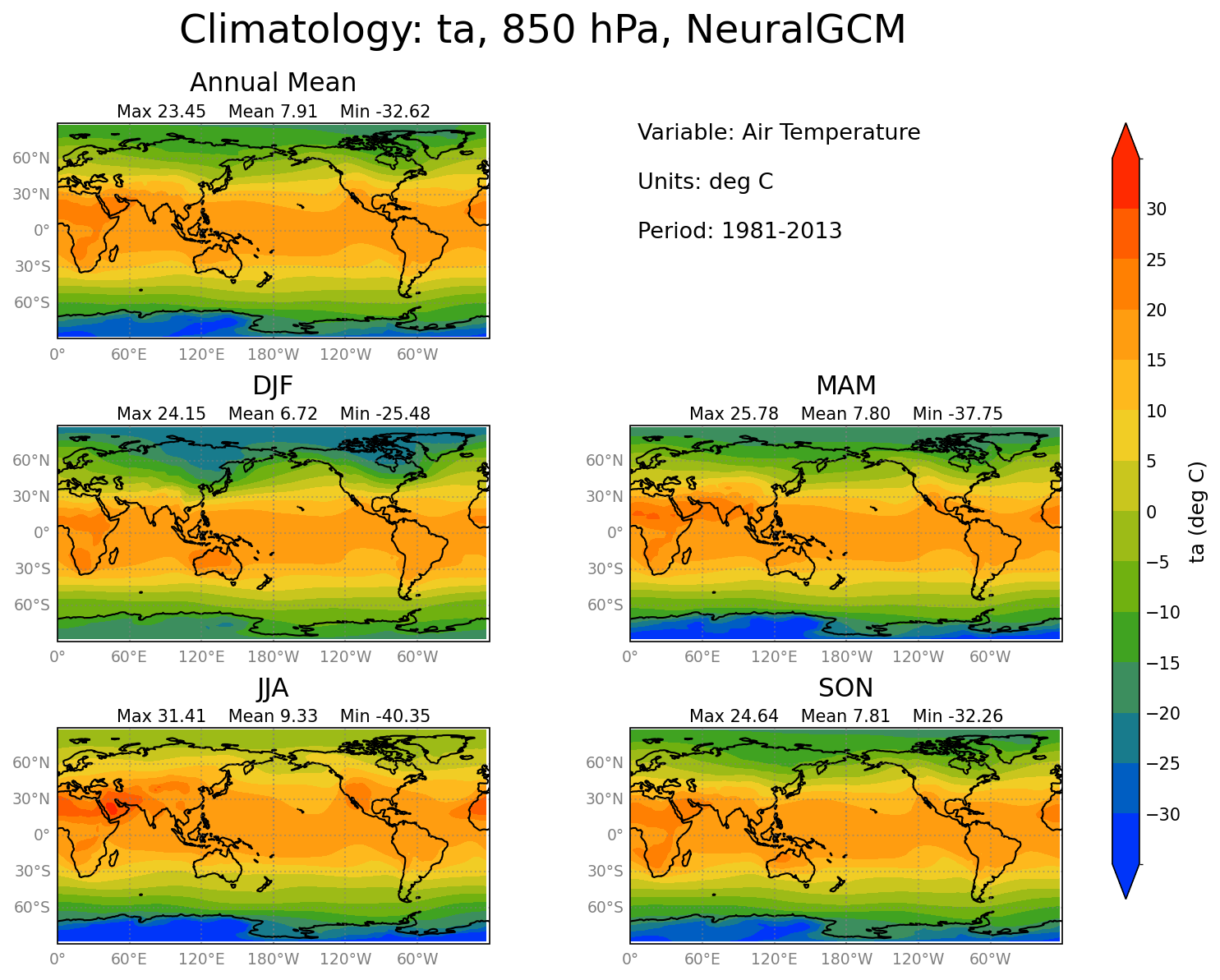}
    \caption{Mean air temperature at 850~hPa (ta-850) simulated by NeuralGCM for 1981--2013. The annual and seasonal means are shown in the same panel layout as Fig.~\ref{SIfig:clim_ACE_ta200}.}
    \label{SIfig:clim_NeuralGCM_ta850}
\end{figure}

%---------------------- FIGURE S11 --------------------------
\begin{figure}[h]
    \centering
    \includegraphics[width=0.5\textwidth]{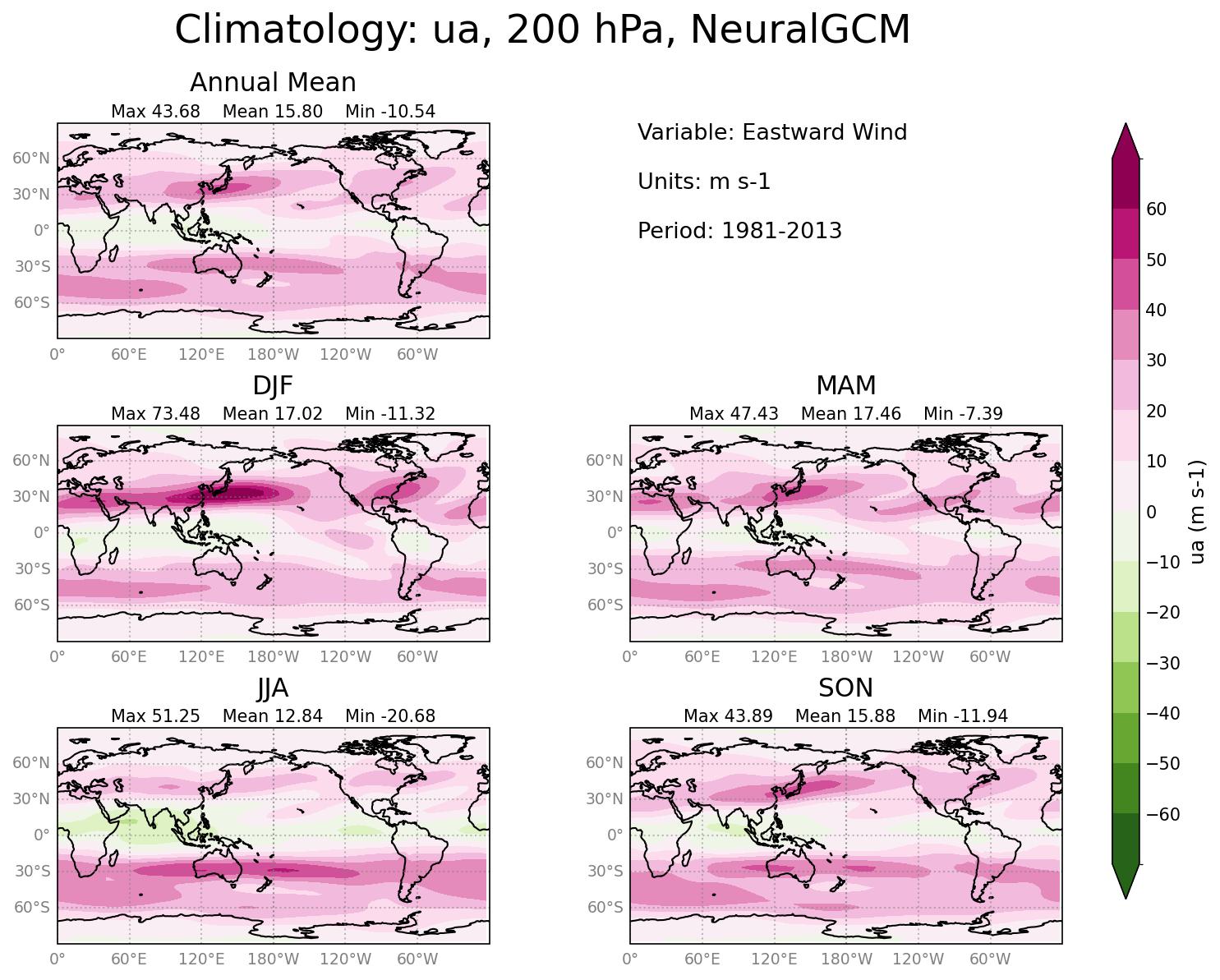}
    \caption{Mean zonal wind at 200~hPa (ua-200) simulated by NeuralGCM for 1981--2013. The annual and seasonal means are shown in the same panel layout as Fig.~\ref{SIfig:clim_ACE_ta200}.}
    \label{SIfig:clim_NeuralGCM_ua200}
\end{figure}

%---------------------- FIGURE S12 --------------------------
\begin{figure}[h]
    \centering
    \includegraphics[width=0.5\textwidth]{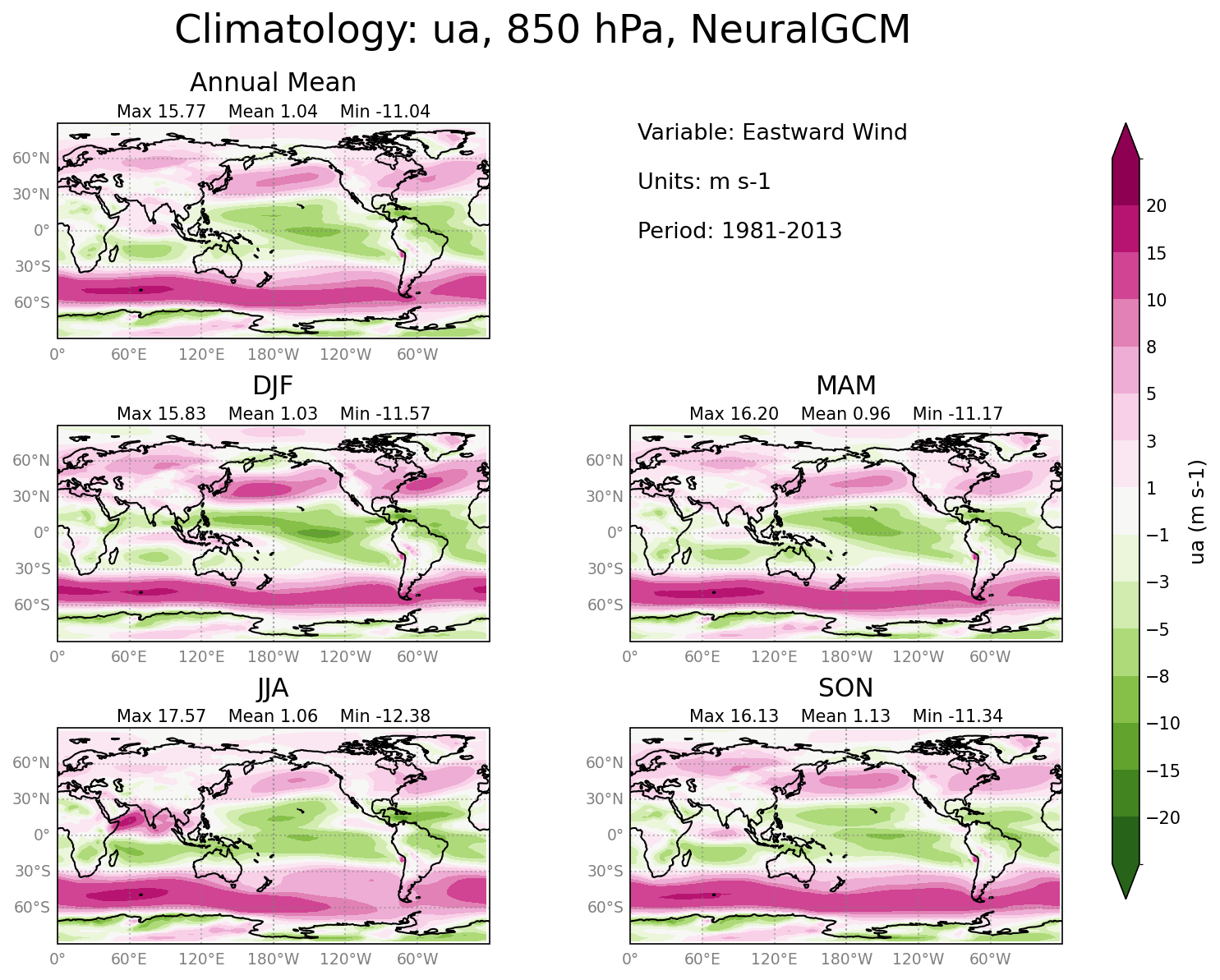}
    \caption{Mean zonal wind at 850~hPa (ua-850) simulated by NeuralGCM for 1981--2013. The annual and seasonal means are shown in the same panel layout as Fig.~\ref{SIfig:clim_ACE_ta200}.}
    \label{SIfig:clim_NeuralGCM_ua850}
\end{figure}

%---------------------- FIGURE S13 --------------------------
\begin{figure}[h]
    \centering
    \includegraphics[width=0.5\textwidth]{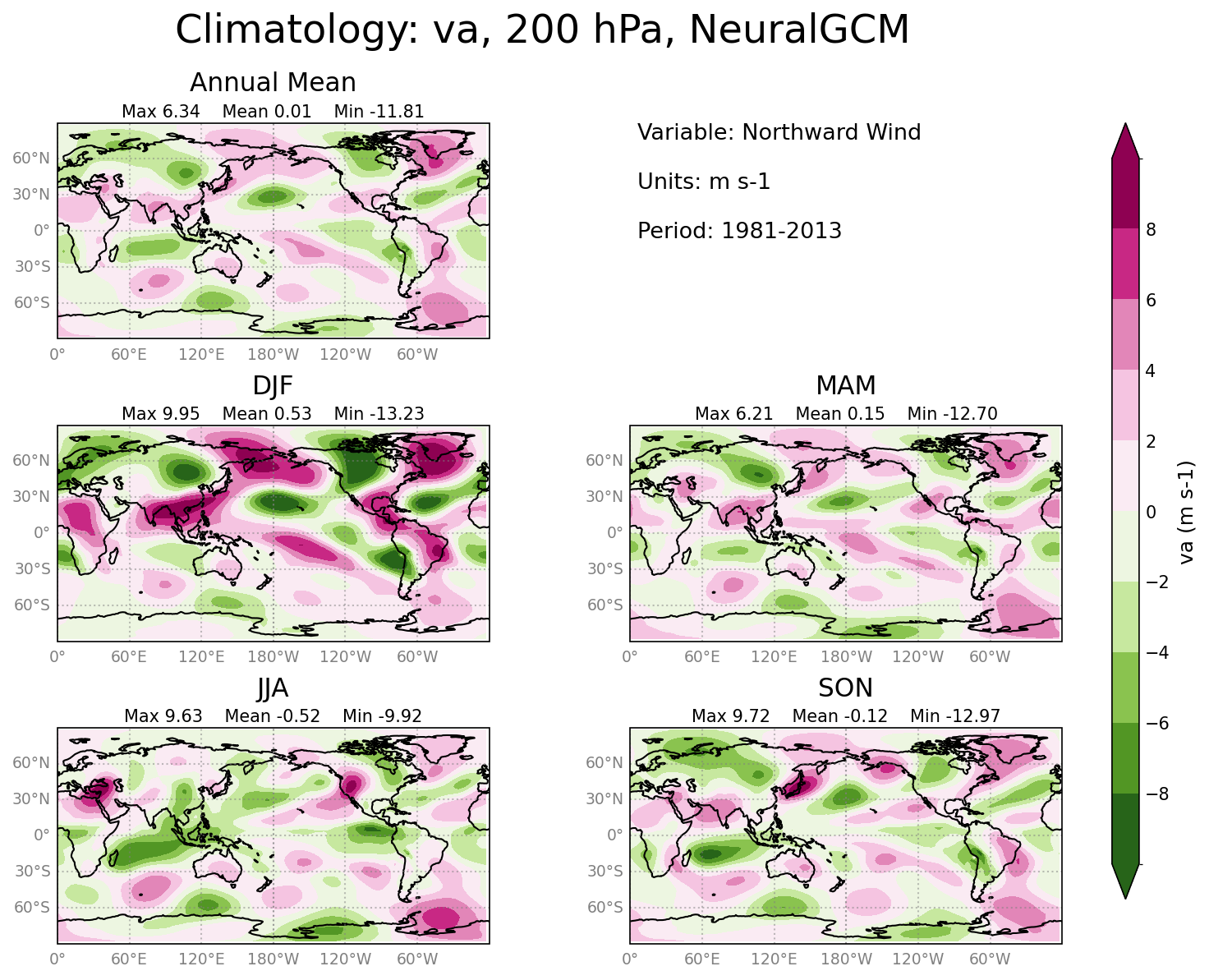}
    \caption{Mean meridional wind at 200~hPa (va-200) simulated by NeuralGCM for 1981--2013. The annual and seasonal means are shown in the same panel layout as Fig.~\ref{SIfig:clim_ACE_ta200}.}
    \label{SIfig:clim_NeuralGCM_va200}
\end{figure}

%---------------------- FIGURE S14 --------------------------
\begin{figure}[h]
    \centering
    \includegraphics[width=0.5\textwidth]{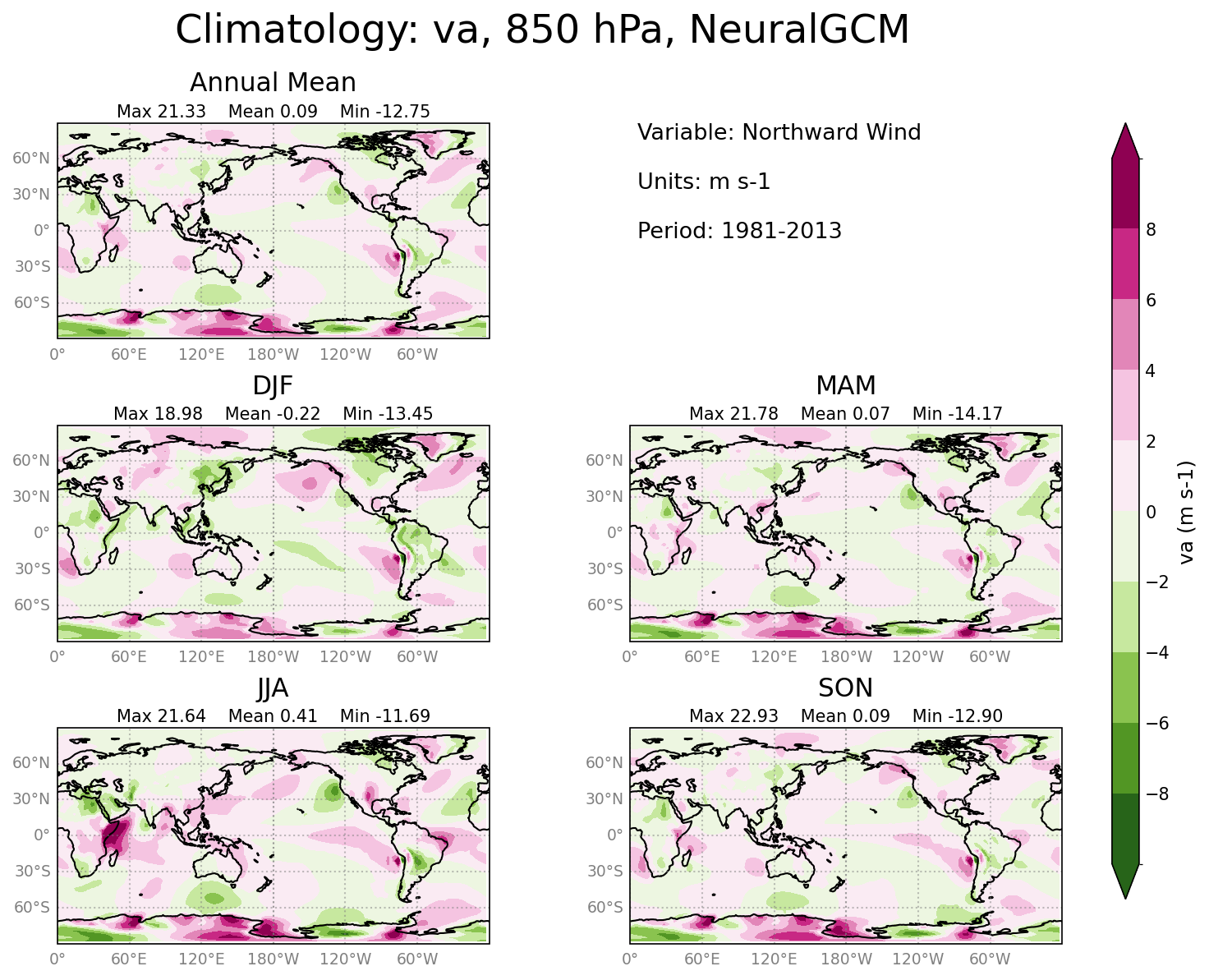}
    \caption{Mean meridional wind at 850~hPa (va-850) simulated by NeuralGCM for 1981--2013. The annual and seasonal means are shown in the same panel layout as Fig.~\ref{SIfig:clim_ACE_ta200}.}
    \label{SIfig:clim_NeuralGCM_va850}
\end{figure}

%---------------------- FIGURE S15 --------------------------
\begin{figure}[h]
    \centering
    \includegraphics[width=0.5\textwidth]{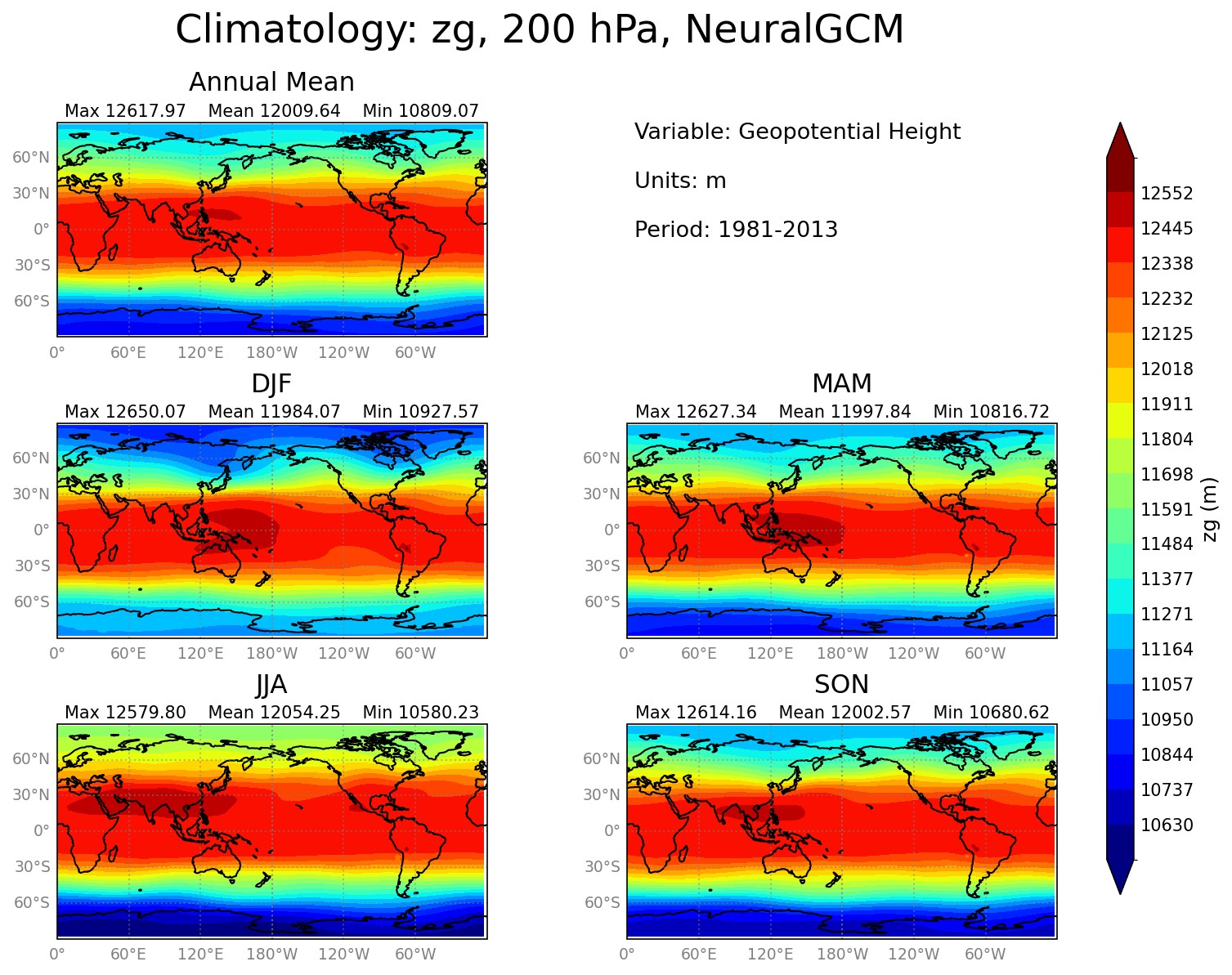}
    \caption{Mean geopotential height at 200~hPa (zg-200) simulated by NeuralGCM for 1981--2013. The annual and seasonal means are shown in the same panel layout as Fig.~\ref{SIfig:clim_ACE_ta200}.}
    \label{SIfig:clim_NeuralGCM_va850}
\end{figure}

%---------------------- FIGURE S16 --------------------------
\begin{figure}[h]
    \centering
    \includegraphics[width=0.5\textwidth]{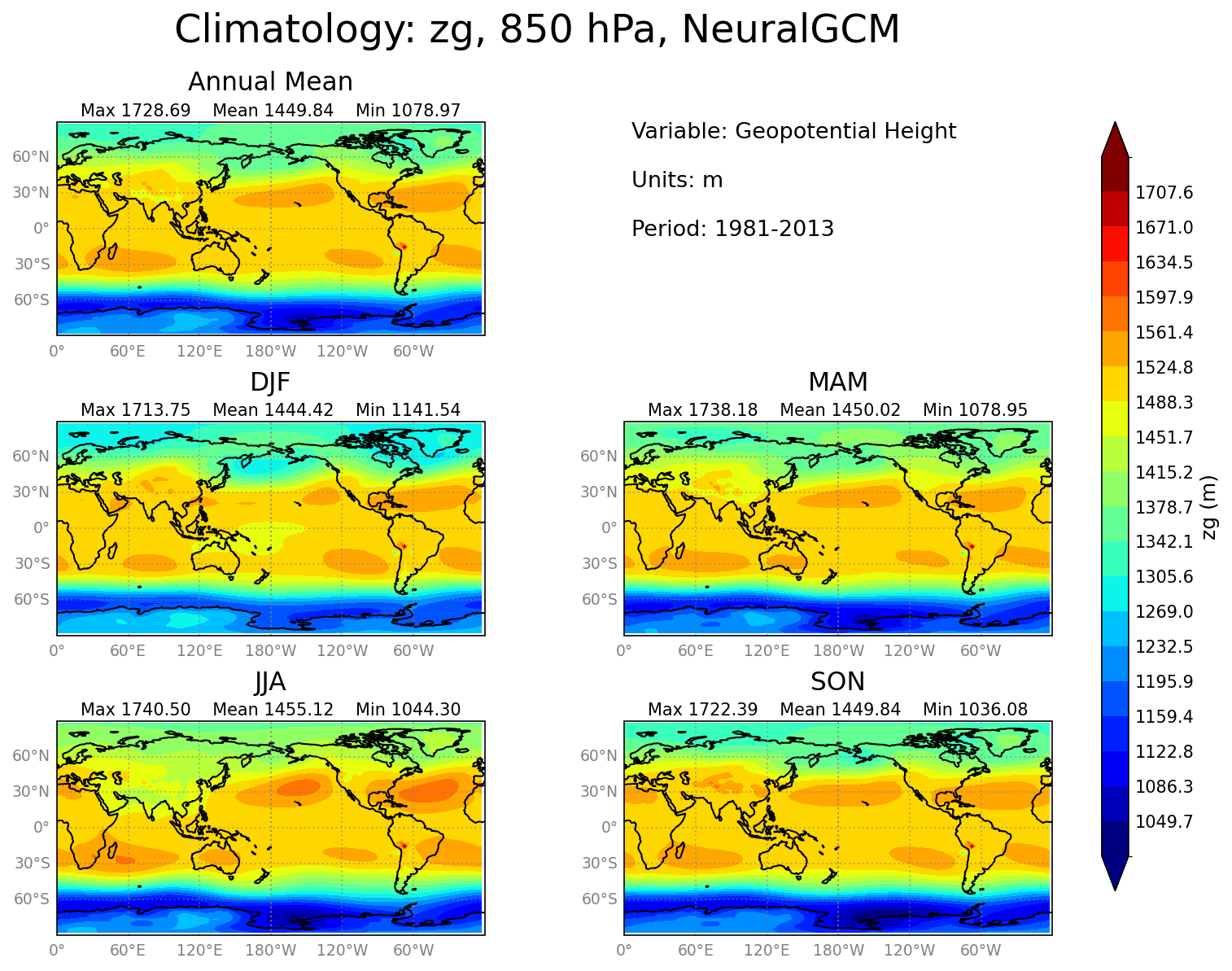}
    \caption{Mean geopotential height at 850~hPa (zg-850) simulated by NeuralGCM for 1981--2013. The annual and seasonal means are shown in the same panel layout as Fig.~\ref{SIfig:clim_ACE_ta200}.}
    \label{SIfig:clim_NeuralGCM_va850}
\end{figure}
%---------------------- FIGURE S17 --------------------------
\begin{figure}[h]
    \centering
    \includegraphics[width=0.5\textwidth]{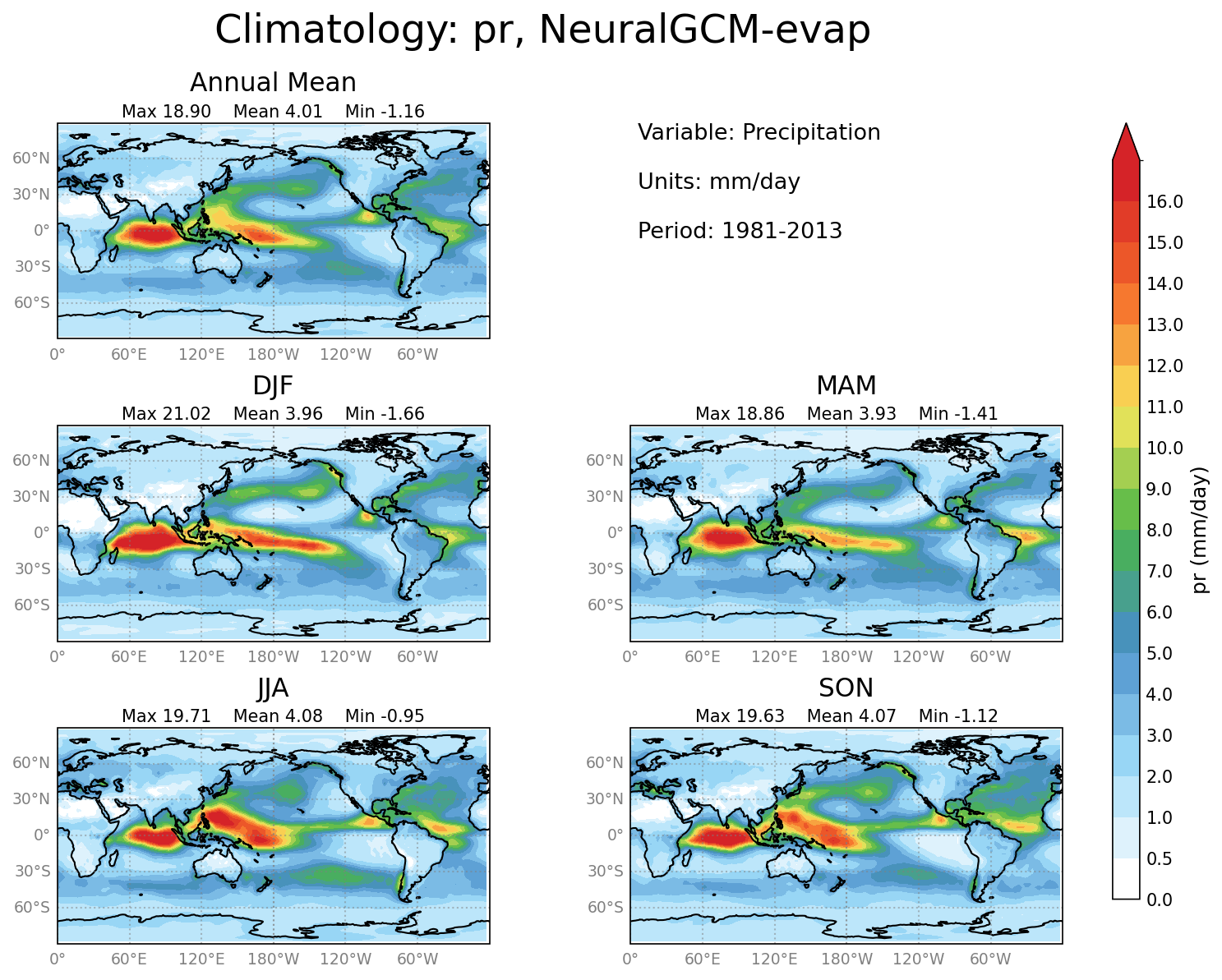}
    \caption{Mean precipitation (pr) simulated by NeuralGCM-evap for 1981--2013. The annual and seasonal means are shown in the same panel layout as Fig.~\ref{SIfig:clim_ACE_ta200}.}
    \label{SIfig:clim_NeuralGCM_zg850}
\end{figure}

%---------------------- FIGURE S18 --------------------------
\begin{figure}[h]
    \centering
    \includegraphics[width=0.5\textwidth]{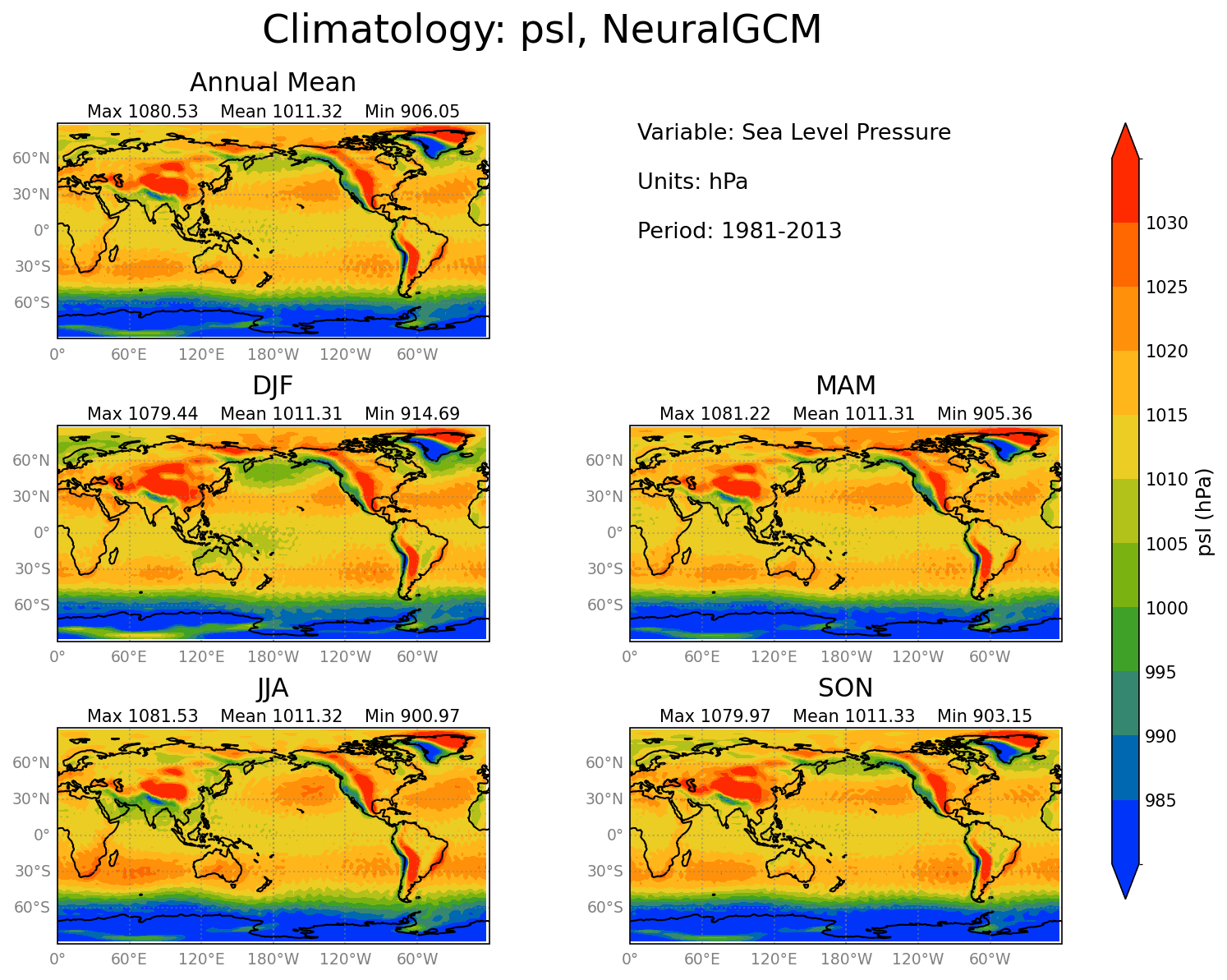}
    \caption{Mean sea-level pressure (psl) simulated by NeuralGCM for 1981--2013. The annual and seasonal means are shown in the same panel layout as Fig.~\ref{SIfig:clim_ACE_ta200}.}
    \label{SIfig:clim_NeuralGCM_psl}
\end{figure}

%---------------------- FIGURE S19 --------------------------
\begin{figure}[h]
    \centering
    \includegraphics[width=0.5\textwidth]{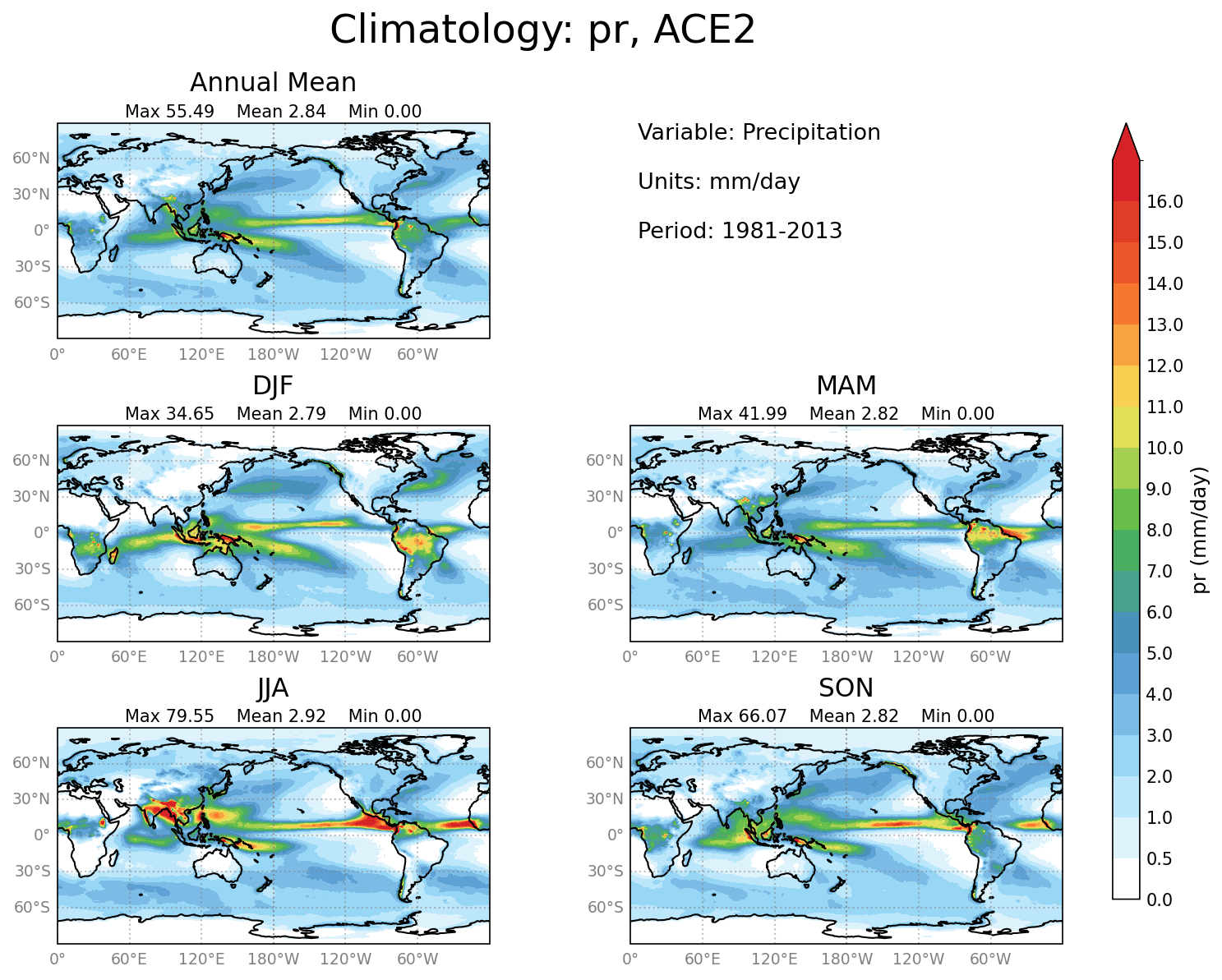}
    \caption{Mean precipitation (pr) simulated by ACE2 for 1981--2013. The annual and seasonal means are shown in the same panel layout as Fig.~\ref{SIfig:clim_ACE_ta200}.}
    \label{SIfig:clim_ACE_pr}
\end{figure}
%---------------------- FIGURE S20 --------------------------

%---------------------- FIGURE S21 --------------------------
\begin{figure}[h]
    \centering
    \includegraphics[width=1\textwidth]{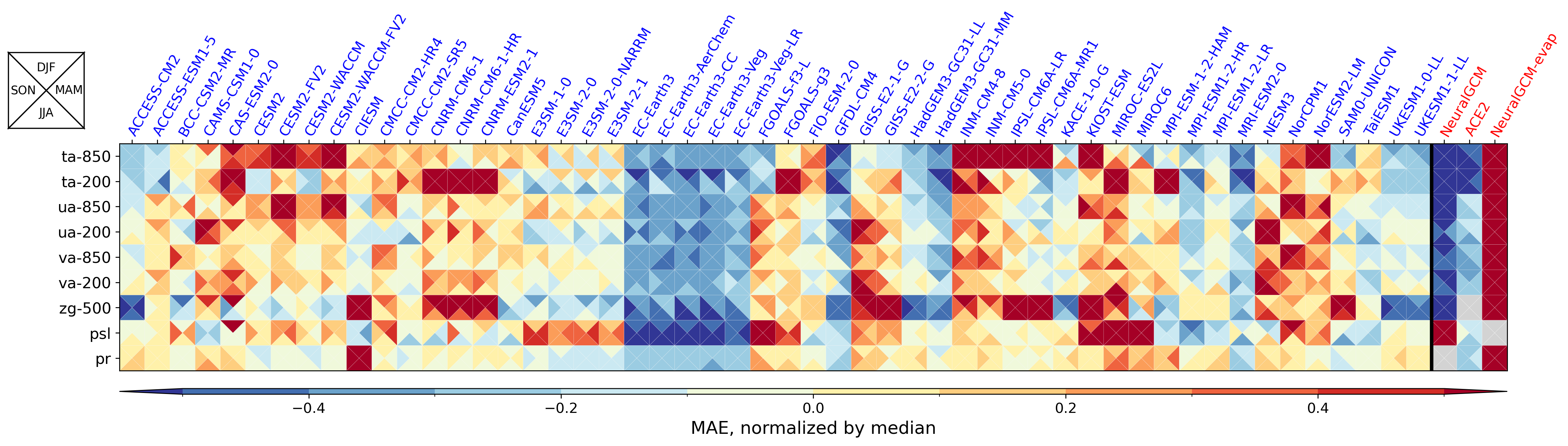}
    \caption{Portrait plot for seasonal mean absolute error (MAE). Values are normalized by the multi-model median following the PMP convention; negative values indicate lower error than the median. Each triangular sector corresponds to a season, as indicated by the legend. The metric is computed against the reference dataset associated with each variable in Table~\ref{tab:var}. Gray cells denote variables unavailable from a given model configuration.}
    \label{SIfig:portrait_MAE}
\end{figure}

\clearpage
\subsection{Madden--Julian Oscillation diagnostics}
\label{SI:MJO}

The MJO diagnostics shown here complement the boreal-winter results in the main manuscript. The wavenumber--frequency spectra report the May--October season, and the companion EWR bar chart summarizes model and ensemble EWR values against the GPCP reference using the same graphical convention as the main text. The east--west power ratio (EWR) compares eastward-propagating power in the MJO band with the corresponding westward-propagating power. Values larger than one indicate a dominance of eastward propagation, while values closer to the observational reference indicate a more realistic propagation amplitude.

%---------------------- FIGURE S22 --------------------------
\begin{figure}[h]
    \centering
    \includegraphics[width=1\textwidth]{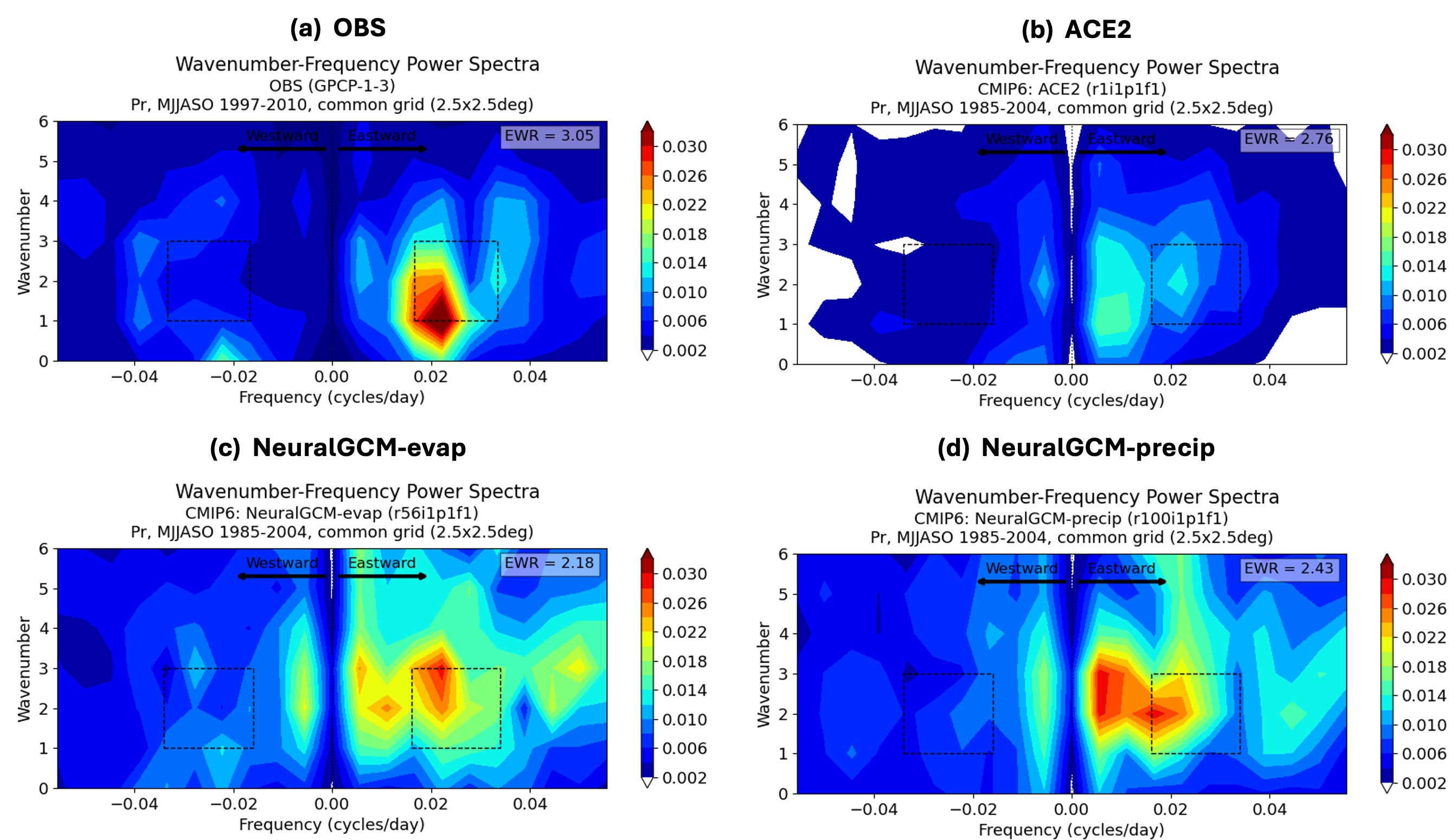}
    \caption{MJO propagation diagnostics for the May--October season. Wavenumber--frequency power spectra of equatorially averaged precipitation ($10^\circ$S--$10^\circ$N) are shown for (a) GPCP v1.3 observations \citep{Huffman2001}, (b) ACE2, (c) NeuralGCM-evap, and (d) NeuralGCM-precip. Dashed boxes mark the westward- and eastward-propagating MJO bands used to compute the EWR. The spectra are averaged over all available years for each dataset; power units are mm$^2$~day$^{-2}$ per frequency interval per wavenumber.}
    \label{SIfig:MJOprop_summer}
\end{figure}

%---------------------- FIGURE S23 --------------------------
% NOTE: The original source used Paper_figures/MJO_NDJFMA.png here. If this supplementary figure is intended to show the May--October EWR ensemble summary, replace this file with the corresponding May--October graphic before final compilation.
\begin{figure}[h]
    \centering
    \includegraphics[width=1\textwidth]{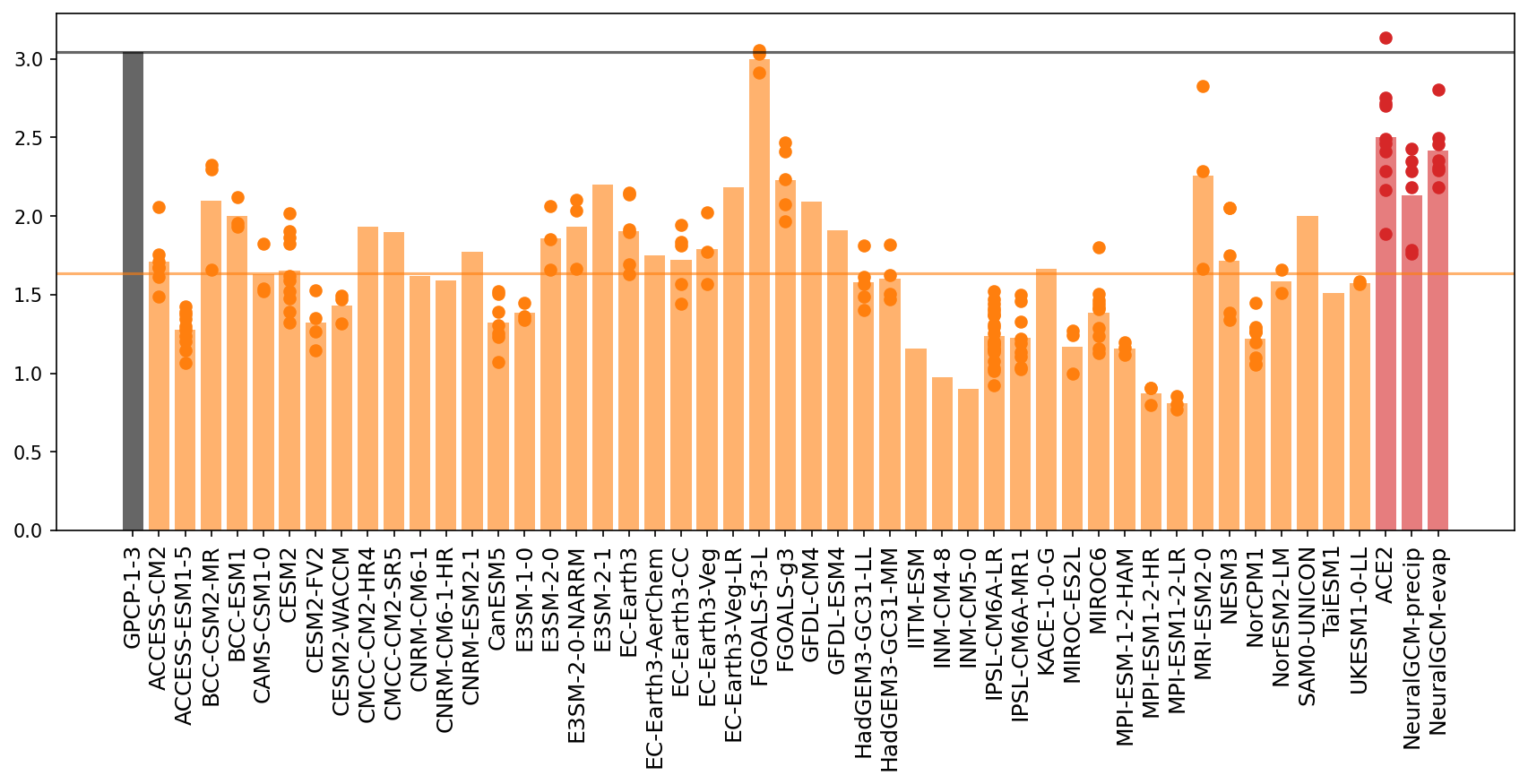}
    \caption{MJO east--west power ratio (EWR; unitless) for CMIP6 models and DL-ESMs. Orange bars show CMIP6 model means, red bars show DL-ESM means, and dots indicate individual ensemble members where available. The gray bar and black horizontal line denote the GPCP v1.3 observational reference \citep{Huffman2001}. Ensemble members and sample sizes follow the convention used for Fig.~\ref{fig:portrait_MoV_ampl}.}
    \label{SIfig:MJO_ensembles_summer}
\end{figure}

\clearpage
\subsection{Regional monsoon diagnostics}
\label{SI:monsoon}

The regional monsoon diagnostic evaluates the seasonal timing of precipitation accumulation in six monsoon domains. The cumulative pentad fraction highlights phase errors in the simulated seasonal cycle, while the onset and decay markers summarize timing differences between each model and the observational reference.

%---------------------- FIGURE S24 --------------------------
\begin{figure}[h]
    \centering
    \includegraphics[width=0.7\textwidth]{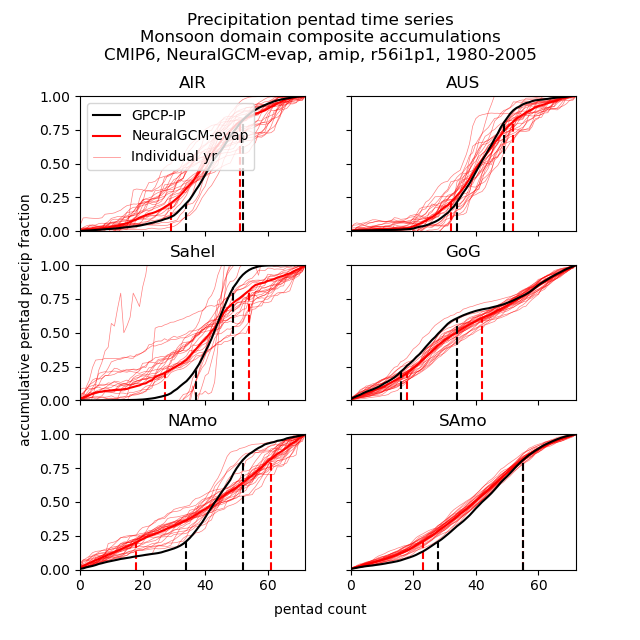}
    \caption{Cumulative pentad precipitation fractions for NeuralGCM-evap and the observational reference over six monsoon regions for 1980--2005: all-India rainfall (AIR), northern Australia (AUS), Sahel, Gulf of Guinea (GoG), North American monsoon (NAMo), and South American monsoon (SAMo). The red line shows the NeuralGCM-evap mean, thin red lines show individual years, and the black line shows the observational reference. Vertical dashed lines indicate the estimated monsoon onset and decay for the observation (black) and the model (red).}
    \label{SIfig:monsoon_Sper_evap}
\end{figure}

\clearpage
\subsection{Precipitation variability diagnostics}
\label{SI:precip_variability}

The precipitation variability maps extend the annual-scale results shown in the main manuscript to semi-annual and interannual timescales. In each panel, the value in parentheses is the PMP spatial-standard-deviation ratio relative to the observational reference. A value of one indicates observed-amplitude variability, while values above or below one indicate overestimated or underestimated variability amplitude, respectively.

%---------------------- FIGURE S25 --------------------------
\begin{figure}[h]
    \centering
    \includegraphics[width=0.9\textwidth]{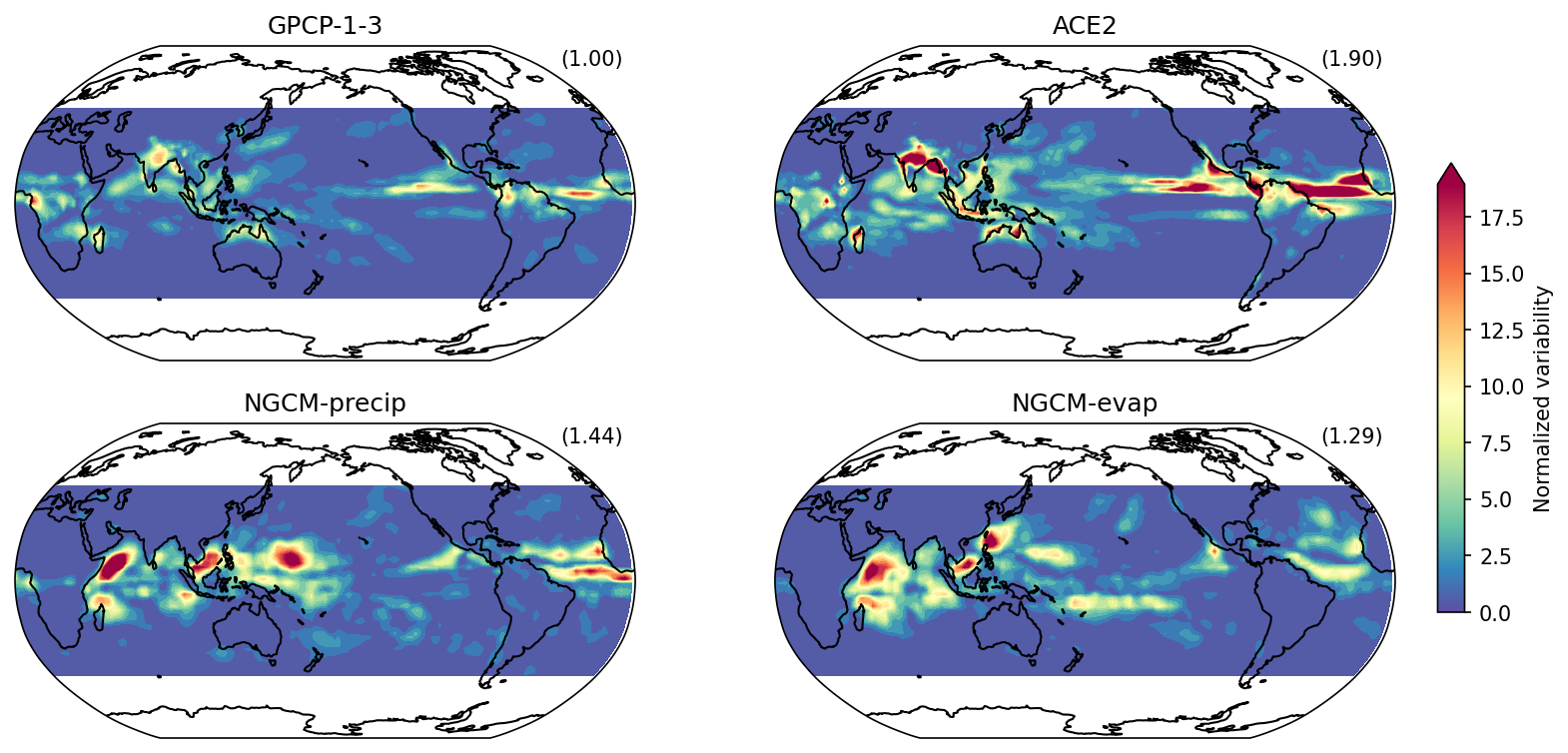}
    \caption{Normalized semi-annual precipitation variability for GPCP v1.3, ACE2, NeuralGCM-precip, and NeuralGCM-evap. The value in parentheses is the PMP spatial-standard-deviation ratio relative to the observed field.}
    \label{SIfig:precip_semiannual}
\end{figure}

%---------------------- FIGURE S26 --------------------------
\begin{figure}[h]
    \centering
    \includegraphics[width=0.9\textwidth]{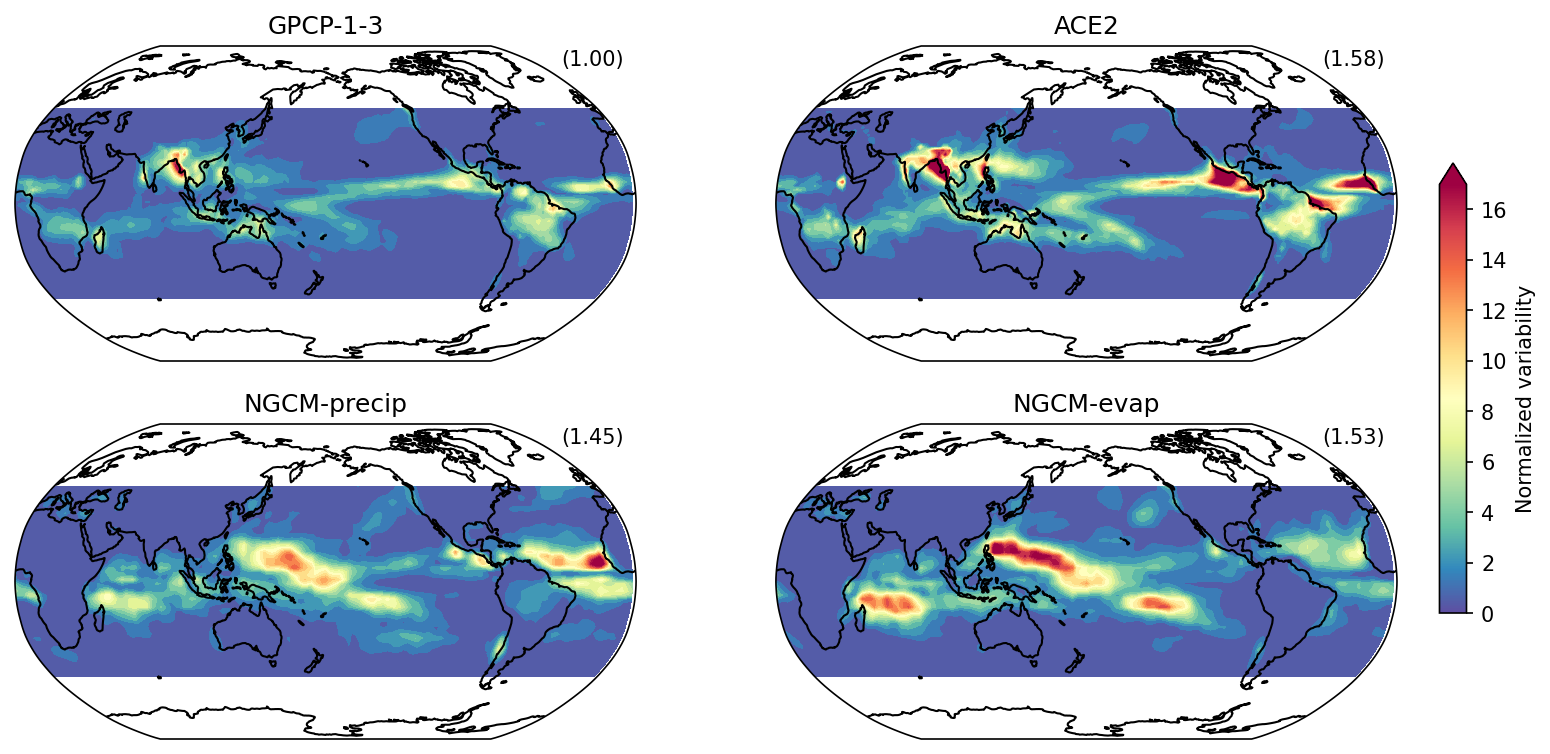}
    \caption{Normalized interannual precipitation variability for GPCP v1.3, ACE2, NeuralGCM-precip, and NeuralGCM-evap. The value in parentheses is the PMP spatial-standard-deviation ratio relative to the observed field.}
    \label{SIfig:precip_inter}
\end{figure}